\documentclass[10pt,aps,prd,floats,floatfix,superscriptaddress,twocolumn,nofootinbib,amsmath,amssymb,showpacs,showkeys]{revtex4-2}

\usepackage[dvipsnames]{xcolor}
\usepackage{multirow}
\usepackage{graphicx}
\usepackage{hyperref}
\hypersetup{colorlinks=true, linkcolor=black, citecolor=black, urlcolor=black}
\usepackage{verbatim}
\usepackage{indentfirst}
\usepackage[normalem]{ulem}
\usepackage{appendix}
\usepackage{soul}
\usepackage{bm}
\usepackage{float}
\usepackage{tikz}
\usepackage{mathtools}
\usepackage{titlesec}
\usepackage{wrapfig}
\usepackage{setspace}
\usepackage{ragged2e}
\usepackage{afterpage}
\usepackage{subcaption}
\usepackage{tablefootnote}
\usepackage{threeparttable}
\usepackage{comment}

\allowdisplaybreaks[2]

\begin{document}

\title{Inflation without an Inflaton III: non-Gaussian signatures}

\author{Mariam Abdelaziz}
\thanks{mariamtarekmohamed.abdelaziz-ssm@unina.it}
\affiliation{Scuola Superiore Meridionale, Largo San Marcellino 10, I-80138 Napoli, Italy}
\affiliation{INFN, Sezione di Napoli, Via Cinthia Edificio 6, I-80126 Napoli, Italy}
\affiliation{Department of Astronomy, Space Science and Meteorology, Cairo University, 12613 Giza, Egypt}

\author{Marisol Traforetti}
\thanks{marisol.traforetti@icc.ub.edu}
\affiliation{ICC, University of Barcelona, Mart\'i i Franqu\`es, 1, E08028 Barcelona, Spain}

\author{Daniele Bertacca}
\thanks{ daniele.bertacca@pd.infn.it}
\affiliation{Dipartimento di Fisica e Astronomia Galileo Galilei, Universit\`a degli Studi di Padova, via Marzolo 8, I-35131, Padova, Italy}
\affiliation{INFN, Sezione di Padova, via Marzolo 8, I-35131, Padova, Italy}
\affiliation{INAF-Osservatorio Astronomico di Padova, Vicolo dell'Osservatorio 5, I-35122 Padova, Italy}

\author{Raul Jimenez}
\thanks{raul.jimenez@icc.ub.edu}
\affiliation{ICC, University of Barcelona, Mart\'i i Franqu\`es, 1, E08028 Barcelona, Spain}
\affiliation{ICREA, Pg. Lluis Companys 23, Barcelona, 08010, Spain}

\author{Sabino Matarrese}
\thanks{ sabino.matarrese@pd.infn.it}
\affiliation{Dipartimento di Fisica e Astronomia Galileo Galilei, Universit\`a degli Studi di Padova, via Marzolo 8, I-35131, Padova, Italy}
\affiliation{INFN, Sezione di Padova, via Marzolo 8, I-35131, Padova, Italy}
\affiliation{INAF-Osservatorio Astronomico di Padova, Vicolo dell'Osservatorio 5, I-35122 Padova, Italy}
\affiliation{Gran Sasso Science Institute, Viale F. Crispi 7, I-67100 L'Aquila, Italy}

\author{Angelo Ricciardone}
\thanks{angelo.ricciardone@unipi.it}
\affiliation{Dipartimento di Fisica ``Enrico Fermi'', Universit\`a di Pisa, Pisa I-56127, Italy}
\affiliation{INFN sezione di Pisa, Pisa I-56127, Italy}

\begin{abstract}
We investigate primordial non-Gaussianity in the Inflation without an Inflaton (IWI) framework, where scalar perturbations are generated at second order by primordial gravitational waves in Einstein gravity on an exact de Sitter (dS) background. Since scalar modes are produced nonlinearly from tensor modes, non-Gaussianity is an intrinsic prediction of the mechanism. We compute the corresponding scalar bispectrum, derive the relevant contribution to the three-point function of the scalar potential, and evaluate its shape numerically. We find that, unlike the scalar power spectrum, the bispectrum depends logarithmically on the ultraviolet cutoff set by the end of inflation, indicating a structural difference between the two- and three-point statistics in this scenario. Its shape is enhanced toward squeezed configurations, but its amplitude becomes strongly suppressed once the scalar power spectrum is normalized to the observed value. The resulting non-Gaussianity at CMB scales is therefore negligibly small, well below present observational sensitivity.
\end{abstract}
\maketitle

\section{INTRODUCTION}
\label{sec:Intro}

The inflationary paradigm was originally introduced to address the shortcomings of the standard Big Bang model by positing a period of accelerated expansion in the early Universe. During this phase, quantum fluctuations are stretched to cosmological scales and provide the initial conditions for the Universe we observe today \cite{Starobinsky:1979ty,Starobinsky:1980te,Guth:1980zm,Linde:1981mu,Mukhanov:1981xt,Albrecht:1982wi,Starobinsky:1982ee,Rubakov:1982df,Guth:1982ec,Linde:1983gd,Kofman:1985aw}. In standard inflationary models, these perturbations originate from quantum fluctuations of a scalar field, the inflaton, whose dynamics drive both the accelerated expansion and the generation of nearly scale-invariant, Gaussian primordial perturbations \cite{Planck:2018jri,Planck:2018vyg}.

Despite its success, the nature of the inflaton and its potential remain unknown. This motivates the exploration of alternative mechanisms in which both the accelerated expansion and the generation of primordial perturbations can be realized without introducing a fundamental scalar degree of freedom.

A concrete realization of this idea is provided by the framework of \emph{Inflation without an Inflaton}, recently developed in Refs.~\cite{Bertacca:2024zfb,Traforetti:2025cax}. In this setup, the background expansion is taken to be exact de Sitter and driven by a cosmological constant, while gravity is described by the Einstein theory. Scalar perturbations are absent at linear order and instead arise dynamically at second order, sourced by tensor perturbations through the nonlinear structure of Einstein's equations. When treated beyond linear order, gravitational waves behave as an effective fluid with nonvanishing energy density, pressure, and anisotropic stress, which in turn source scalar metric perturbations on super-horizon scales. This effective description captures the cumulative contribution of sub-horizon tensor fluctuations generated during inflation.

In Refs.~\cite{Bertacca:2024zfb,Traforetti:2025cax}, it was shown that this mechanism produces a scale-invariant scalar power spectrum consistent with observations. Its amplitude fixes a relation between the inflationary scale $H_{\rm inf}$, the number of observed e-folds $N_{\rm obs}$, and the predicted tensor-to-scalar ratio. An important outcome of that analysis is the ultraviolet sensitivity of the scalar power spectrum, which ties the amplitude of scalar perturbations to the end of inflation and connects naturally to the notion of quantum break time and to the number of particle species \cite{Dvali:2017eba}.

While the power spectrum captures the leading statistical properties of primordial perturbations, higher-order correlation functions provide a more sensitive probe of the mechanism responsible for their generation. Primordial non-Gaussianity, in particular, encodes information about nonlinear interactions and is widely regarded as a powerful discriminator among competing inflationary scenarios \cite{Maldacena:2002vr,Bartolo:2004if}. In standard single-field slow-roll inflation, the bispectrum is typically suppressed by slow-roll parameters \cite{Maldacena:2002vr,Creminelli:2004yq}. Departures from this minimal framework, such as non-standard initial states, additional light degrees of freedom, or purely gravitational sources, can instead lead to distinctive non-Gaussian signatures \cite{Bartolo:2001cw,Chen:2006nt,Planck:2018jri,Planck:2019kim,Planck:2015zfm,Planck:2013wtn}.

In the IWI scenario, scalar perturbations arise only at second order and are sourced by tensor perturbations. Even if primordial gravitational waves are Gaussian, the quadratic nature of their contribution to the scalar sector implies that non-Gaussianity is unavoidably generated. The scalar bispectrum is therefore a robust prediction of the IWI framework, arising purely from gravitational nonlinearities rather than from self-interactions of scalar fields.

Tensor-induced non-Gaussianity has also been investigated within the standard inflationary framework, where gravitational waves source scalar fluctuations at second order and contribute to higher-order correlation functions \cite{Abdelaziz:2025qpn}. However, those scenarios typically rely on the presence of an underlying inflaton field. In contrast, the IWI framework relies exclusively on pure Einstein gravity on an exact dS background, isolating the purely gravitational nonlinearities without invoking additional degrees of freedom.

In this paper, we compute the three-point function of the scalar gravitational potential induced at second order by tensor perturbations in a pure dS background. In Sec.~\ref{sec:background}, we summarize the perturbative setup and review the key results of Refs.~\cite{Bertacca:2024zfb,Traforetti:2025cax}. In Sec.~\ref {sec:bispectrum}, we define the scalar bispectrum and derive its formal expression in momentum space, identifying the kernel that encodes the nonlinear tensor couplings. In Sec.~\ref{sec:Results}, we present the numerical evaluation of the corresponding bispectrum contribution and study its dependence on the ultraviolet cutoff. We conclude in Sec.~\ref{sec:conclusions}, with additional technical details collected in Appendices~\ref{appendixA} and~\ref{sec:kernel}.

%%%%%%%%%%%%%%%%%%%%%%%%%%%%%%%%%%%%%%%%%
\section{BACKGROUND}
\label{sec:background}

In this work, we build on the analysis developed in Refs.~\cite{Bertacca:2024zfb,Traforetti:2025cax}. We adopt the same setup and briefly summarize here the ingredients relevant to the present discussion. We consider a purely dS background within Einstein gravity. The perturbed metric is written as
\begin{eqnarray}
ds^2 &=& a^2(\eta)\bigg\{ -\left(1 + \psi^{(2)} \right)d\eta^2 \nonumber \\ \quad\quad &&
+\left[\left(1 - \phi^{(2)}\right)\delta_{ij} + \chi^{(1)}_{ij} \right]dx^i dx^j \bigg\}\;,
\end{eqnarray}
where $\chi^{(1)}_{ij}$ denotes the first-order tensor modes, while $\psi^{(2)}$ and $\phi^{(2)}$ are the second-order scalar perturbations.

On the right-hand side of Einstein's field equations, we include the energy-momentum tensor of an effective fluid. This contribution arises from the vacuum expectation value of the second-order Einstein tensor sourced by tensor perturbations \cite{Bertacca:2024zfb}. On sub-horizon scales, gravitational waves behave as an effective fluid with nonvanishing energy density $\rho$, pressure $p$, and anisotropic stress $\pi^\mu_{\ \nu}$. The corresponding energy-momentum tensor takes the form
\begin{eqnarray}
T^\mu_{\ \nu} = (\rho + p) u^\mu u_\nu + p\, \delta^\mu_{\ \nu} + \pi^\mu_{\ \nu}\;,
\end{eqnarray}
where $u^\mu$ is the fluid four-velocity, normalized as $u^\mu u_\mu =~-1$. A cosmological constant $\Lambda$ is also included, since it drives the dS expansion. We consider generic $w$ and $c_s$, where $w=\Bar{p}/\Bar{\rho}$ is the background equation-of-state parameter and $c_s^2=p^{(2)}/\rho^{(2)}$ is the adiabatic sound speed.

Expanding Einstein's equations to quadratic order in $\chi^{(1)}_{ij}$ allows one to determine the evolution of scalar perturbations. In the very large-scale limit, and using the traceless part of the $ij$ components, one finds \cite{Bertacca:2024zfb}
\begin{eqnarray}
\label{eq: phi_second_order}
\psi^{(2)}
= \phi^{(2)} - 8\pi G a^2 \Pi^{(2)} - \frac{{\cal F}_\chi}{4}\;,
\end{eqnarray}
where $\Pi^{(2)}$ denotes the scalar component of the second-order anisotropic stress. The quantity ${\cal F}_\chi$ encodes the nonlinear couplings of the first-order tensor modes and is given by
\begin{eqnarray}
\label{eq:F_chi}
{\cal F}_\chi &=&
4\nabla^{-2}\bigg(
\frac{3}{4}\chi_1^{lk,m}\chi_{1kl,m}
+ \frac{1}{2} \chi_1^{kl}\nabla^2 \chi_{1lk}
- \frac{1}{2} \chi_{1,l}^{km}\chi_{1m,k}^{\,l}
\bigg) \nonumber \\ &&
- 6\nabla^{-4}\partial_i\partial^j {\cal A}^i_{\ j}\;,
\end{eqnarray}
with
\begin{eqnarray}
{\cal A}^i_{\ j} &=&
\frac{1}{2}\chi_1^{lk,i}\chi_{1kl,j}
+ \chi_1^{kl}\chi_{1lk,j}^{\ ,i}
- \chi_1^{kl}\chi_{1l,jk}^{\ \ i}
- \chi_1^{kl}\chi_{1lj,k}^{\ \ i} \nonumber \\
&& + \chi_1^{kl}\chi_{1j,kl}^{\ \ i}
+ \chi_{1,l}^{ik}\chi_{1jk}^{\ ,l}
- \chi_{1,l}^{ki}\chi_{1j,k}^{\ l}\;.
\end{eqnarray}
The equation of motion for the potential $\phi^{(2)}$ can be found in Ref.~\cite{Bertacca:2024zfb}.

In Ref.~\cite{Traforetti:2025cax}, the scalar power spectrum generated within the IWI framework was computed by numerically integrating the full second-order kernel. Matching the observed amplitude of scalar fluctuations at CMB scales fixes the inflationary energy scale $H_{\rm inf}$ as a function of the number of observed e-folds $N_{\rm obs}$. In particular, for $N_{\rm obs}\sim 30$ one finds $H_{\rm inf}\sim 3 \times 10^{13}\,\mathrm{GeV}$, while larger values of $N_{\rm obs}$ correspond to lower inflationary scales. We refer the reader to Ref.~\cite{Traforetti:2025cax} for the details of that calculation.

In the following section, we extend this framework to the computation of the scalar bispectrum sourced by tensor perturbations.

%%%%%%%%%%%%%%%%%%%%%%%%%%%%%%%%%%%%%%%%%
\section{BISPECTRUM}
\label{sec:bispectrum}

Following Refs.~\cite{Bertacca:2024zfb,Traforetti:2025cax}, we focus on the analytic case in which $\psi^{(2)} = 0$ and $\Pi^{(2)} = 0$. From Eq.~\eqref{eq: phi_second_order}, the solution for which $\phi^{(2)}$ remains constant on very large scales is then
\begin{equation}
\label{eq: phi constant solution}
\phi^{(2)} = \frac{1}{4}\,{\cal F}_\chi \;.
\end{equation}
The bispectrum of scalar fluctuations $\phi$ is defined as
\begin{align}
\label{Threepnt}
\bigl\langle \phi(\bm k_1)\,\phi(\bm k_2)\,\phi(\bm k_3)\bigr\rangle
&= \frac{1}{8}\,
\bigl\langle \phi^{(2)}(\bm k_1)\,\phi^{(2)}(\bm k_2)\,\phi^{(2)}(\bm k_3)\bigr\rangle
\nonumber\\
&= (2\pi)^3\,\delta^{(3)}(\bm k_1+\bm k_2+\bm k_3)
\nonumber\\
&\qquad  B_{\phi}(k_1,k_2,k_3)\; .
\end{align}

In the following, we compute explicitly the contribution arising from the first term in Eq.~\eqref{eq:F_chi}, denoted as the ``$A$'' term in Ref.~\cite{Traforetti:2025cax}. This sector is sufficient to establish the existence of a non-vanishing tensor-induced scalar bispectrum, its logarithmic ultraviolet sensitivity, and its characteristic dependence on triangle shape. The pure ``$B$'' and the mixed  ``$AB$'' sectors are not included in the present numerical analysis, though they are not expected to modify these qualitative features. The explicit Wick contraction of the three-point function in Eq.~\eqref{Threepnt} is presented in Appendix~\ref{appendixA}.
The Dirac delta functions arising from the Wick contractions allow us to perform the integrals over $\bm p_2$ and $\bm p_3$, leaving a single integral over $\bm p_1$. By comparing Eq.~\eqref{Threepnt} with Eq.~\eqref{finalthreepnt}, the bispectrum can be written as
\begin{align}
B_\phi(k_1,k_2,k_3)
&= \frac{\pi^3}{k_1^2 k_2^2 k_3^2}
\int d^3 \bm{p}_1\;
K(\bm{p}_1;\bm{k}_1,\bm{k}_2,\bm{k}_3)
\nonumber\\
&\quad\times
\frac{\Delta^2_\chi(p_1)}{p_1^3}
\frac{\Delta^2_\chi(p_2)}{p_2^3}
\frac{\Delta^2_\chi(p_3)}{p_3^3}\;,
\end{align}
where the kernel $K(\bm{p}_1;\bm{k}_1,\bm{k}_2,\bm{k}_3)$ is defined in Eq.~\eqref{eq: explicit kernel K}, including the sum over the polarization indices, and is computed explicitly in Appendix~\ref{sec:kernel}.

On super-horizon scales in dS space, the dimensionless tensor power spectrum is scale invariant and given by
\begin{equation}
\label{TensorPS}
\Delta_h^2(k) = \frac{16}{\pi}
\left( \frac{H_{\rm inf}}{m_{\rm pl}} \right)^2\;,
\end{equation}
where $H_{\rm inf}=\sqrt{\Lambda/3}$ is the Hubble constant during inflation and $m_{\rm pl}=G^{-1/2}$ is the Planck mass. Equation~\eqref{TensorPS} already accounts for the two tensor polarizations, such that $\Delta_\chi^2 = \tfrac{1}{2}\Delta_h^2$. Using Eq.~\eqref{TensorPS}, the bispectrum becomes
\begin{align}
\label{integration of B}
B_\phi(k_1,k_2,k_3)
&= \frac{8^3}{k_1^2 k_2^2 k_3^2}
\left( \frac{H_{\rm inf}}{m_{\rm pl}} \right)^6
\nonumber\\
&\quad \times \int d^3 \bm{p}_1\;
\frac{K(\bm{p}_1;\bm{k}_1,\bm{k}_2,\bm{k}_3)}
{p_1^3 p_2^3 p_3^3}\; .
\end{align}
Once the integral is evaluated numerically, the bispectrum is directly obtained in units of $(H_{\rm inf}/m_{\rm pl})^6$. This parallels the result of Ref.~\cite{Traforetti:2025cax} for the power spectrum, which is expressed in units of $(H_{\rm inf}/m_{\rm pl})^4$.

Before proceeding to the numerical evaluation of the bispectrum, we first examine the ultraviolet behavior of the integral in Eq.~\eqref{integration of B}.

%%%%%%%%%%%%%%%%%%%%%%%%%%%%%
\subsection{Ultraviolet behavior of the bispectrum}

To analyze the ultraviolet behavior of the bispectrum integral in Eq.~\eqref{integration of B}, we first isolate the integrand,
\begin{equation}
I(\bm{p}_1;\bm{k}_1,\bm{k}_2,\bm{k}_3)
\equiv
\frac{K(\bm{p}_1;\bm{k}_1,\bm{k}_2,\bm{k}_3)}
{p_1^3 p_2^3 p_3^3}\,,
\label{eq:definitionI}
\end{equation}
where the explicit expression for the kernel $K(\bm{p}_1;\bm{k}_1,\bm{k}_2,\bm{k}_3)$ is given in Eq.~\eqref{eq: kernel K explicitly}.

In the deep UV regime, $p_1 \gg k_i$, all internal momenta scale uniformly, so that $p_1 \sim p_2 \sim p_3 \sim p$. The angular structures $\mathcal{I}_\alpha$ in Eq.~\eqref{eq: kernel K explicitly} depend only on relative angles and therefore remain $\mathcal{O}(1)$. The tensor contractions entering $K$ then scale homogeneously as $p^6$, implying
\begin{equation}
K(\bm{p}_1;\bm{k}_1,\bm{k}_2,\bm{k}_3) \sim p^6 \, .
\end{equation}
Since the denominator scales as $p_1^3 p_2^3 p_3^3 \sim p^9$, the integrand behaves asymptotically as
\begin{equation}
I(\bm{p}_1;\bm{k}_1,\bm{k}_2,\bm{k}_3) \sim p^{-3} \, .
\end{equation}

To determine the full UV scaling of the integral, the phase-space measure must also be included. In spherical momentum coordinates,
\begin{equation}
d^3\bm{p}_1 = p^2 \, dp \, d\Omega \, ,
\end{equation}
so that the radial integrand scales as
\begin{equation}
p^2 I(p) \sim p^{-1} \, .
\end{equation}
The ultraviolet contribution, therefore, takes the form
\begin{equation}
\int_{p_{\rm min}}^{p_{\rm max}} dp\, p^{-1}
=
\ln\!\left(\frac{p_{\rm max}}{p_{\rm min}}\right) \, ,
\end{equation}
showing that the bispectrum is logarithmically sensitive to the ultraviolet cutoff.\footnote{A straightforward power-counting argument shows that the mixed $AB$ terms, as well as the pure $BB$ contributions, do not modify the logarithmic ultraviolet behavior of the bispectrum.}

To perform this integral, we impose a physical ultraviolet cutoff at the maximum comoving wavenumber that exits the horizon at the end of inflation,
\begin{equation}
p_{\rm max} \equiv k_{\rm end},
\end{equation}
following the prescription of Refs.~\cite{Bertacca:2024zfb,Traforetti:2025cax}. As in Ref.~\cite{Traforetti:2025cax}, we parametrize this cutoff through
\begin{equation}
\frac{k_{\rm end}}{k_0} = e^{N_{\rm obs}}.
\end{equation}
With this choice, the ultraviolet contribution becomes
\begin{equation}
\ln\!\left(\frac{p_{\rm max}}{p_{\rm min}}\right)
=
\ln\!\left(\frac{e^{N_{\rm obs}}k_0}{p_{\rm min}}\right)
=
N_{\rm obs} + \ln\!\left(\frac{k_0}{p_{\rm min}}\right).
\end{equation}

This shows that the ultraviolet contribution depends linearly on the number of observed e-folds $N_{\rm obs}$ and does not introduce any additional dependence on the external scales $k_i$. We stress, however, that the physical relation between $k_{\rm end}$ and $N_{\rm obs}$ is not one-to-one, since it also depends on the post-inflationary history. In this sense, $N_{\rm obs}$ should be regarded here as a convenient parameterization of the ultraviolet cutoff, following the same convention adopted in Ref.~\cite{Traforetti:2025cax}. This logarithmic ultraviolet behavior, together with the absence of any additional dependence on the external scales $k_i$, is also confirmed by our numerical analysis.
%%%%%%%%%%%%%%%%%%%%%%%%%%%
\section{RESULTS}
\label{sec:Results}

Having established the logarithmic ultraviolet scaling of the bispectrum and the absence of any residual running with the overall observable scale within the adopted cutoff prescription, we now present the numerical evaluation of the scalar bispectrum in the IWI framework.

To characterize the primordial non-Gaussian signal, one must determine both its geometric momentum dependence, encoded in the shape function, and its overall magnitude, quantified by the amplitude parameter $f_{\rm NL}$. We address these two aspects in turn.

To evaluate the integral \eqref{integration of B}, we follow the procedure adopted in Refs.~\cite{Espinosa:2018eve,Abdelaziz:2025qpn} and introduce a cylindrical coordinate system such that
\begin{align*}
\bm{k}_{1}=\left(k_{1x}, k_{1y}, 0\right), \quad
\bm{k}_{2}=\left(k_{2x}, k_{2y}, 0\right), \quad
\bm{k}_{3}=\left(-k_{3x}, 0,0\right),
\end{align*}
and
\begin{align*}
\bm{p}_{1} &= (r \cos \alpha,\, r \sin \alpha,\, \ell), \\
\bm{p}_{2} &= \left(-k_{1x}+r \cos \alpha,\,-k_{1y}+r \sin \alpha,\, \ell\right), \\
\bm{p}_{3} &= \left(-k_{3x}+r \cos \alpha,\, r \sin \alpha,\, \ell\right).
\end{align*}
The variables $\ell$, $r$, and $\alpha$ provide a convenient choice of cylindrical coordinates as integration variables. In this way, the integration over $\bm p_1$ becomes
\begin{equation}
\label{int}
\int \mathrm{d}^{3} \bm p_1 \;\longrightarrow\;
\int_{-\infty}^{+\infty} \mathrm{d} \ell
\int_{0}^{+\infty} r\,\mathrm{d} r
\int_{0}^{2 \pi} \mathrm{d} \alpha\;.
\end{equation}

Using in cylindrical coordinates $\{r,\ell,\alpha\}$, we now proceed to evaluate numerically the integral in Eq.~\eqref{integration of B}, with a sharp ultraviolet cutoff imposed at $p_{\rm max}=k_{\rm end}$. We fix the inflationary scale to
$H_{\rm inf} = 3 \times 10^{13}\,\mathrm{GeV},$
as required to reproduce the observed scalar power spectrum in Ref.~\cite{Traforetti:2025cax}. This value corresponds to $N_{\rm obs}=30$, which in turn fixes the ultraviolet cutoff to
\begin{equation}
p_{\rm max} \equiv k_{\rm end} = e^{N_{\rm obs}} k_0
\sim 10^9\,\mathrm{Mpc}^{-1}.
\end{equation}

We first study the dependence of the bispectrum on the triangle geometry. To this end, we introduce the dimensionless shape function \cite{Espinosa:2018eve}:
\begin{equation}
S_\phi(k_1,k_2,k_3)
=
k_1^2 k_2^2 k_3^2 \,
\frac{B_\phi(k_1,k_2,k_3)}
{\sqrt{\Delta_\phi^2(k_1)\Delta_\phi^2(k_2)\Delta_\phi^2(k_3)}} .
\end{equation}
We evaluate this quantity by fixing one external momentum to the CMB pivot scale,
\begin{equation}
k_3 \equiv k_{\rm CMB} = 0.01\,\mathrm{Mpc}^{-1},
\end{equation}
and scanning the ratios $(k_1/k_3,\,k_2/k_3)$ over the domain allowed by the triangle inequalities. We impose the ordering $k_1 \le k_2 \le k_3$, so that each configuration is counted only once.

\begin{figure}[t!]
\centering
\includegraphics[width=\linewidth]{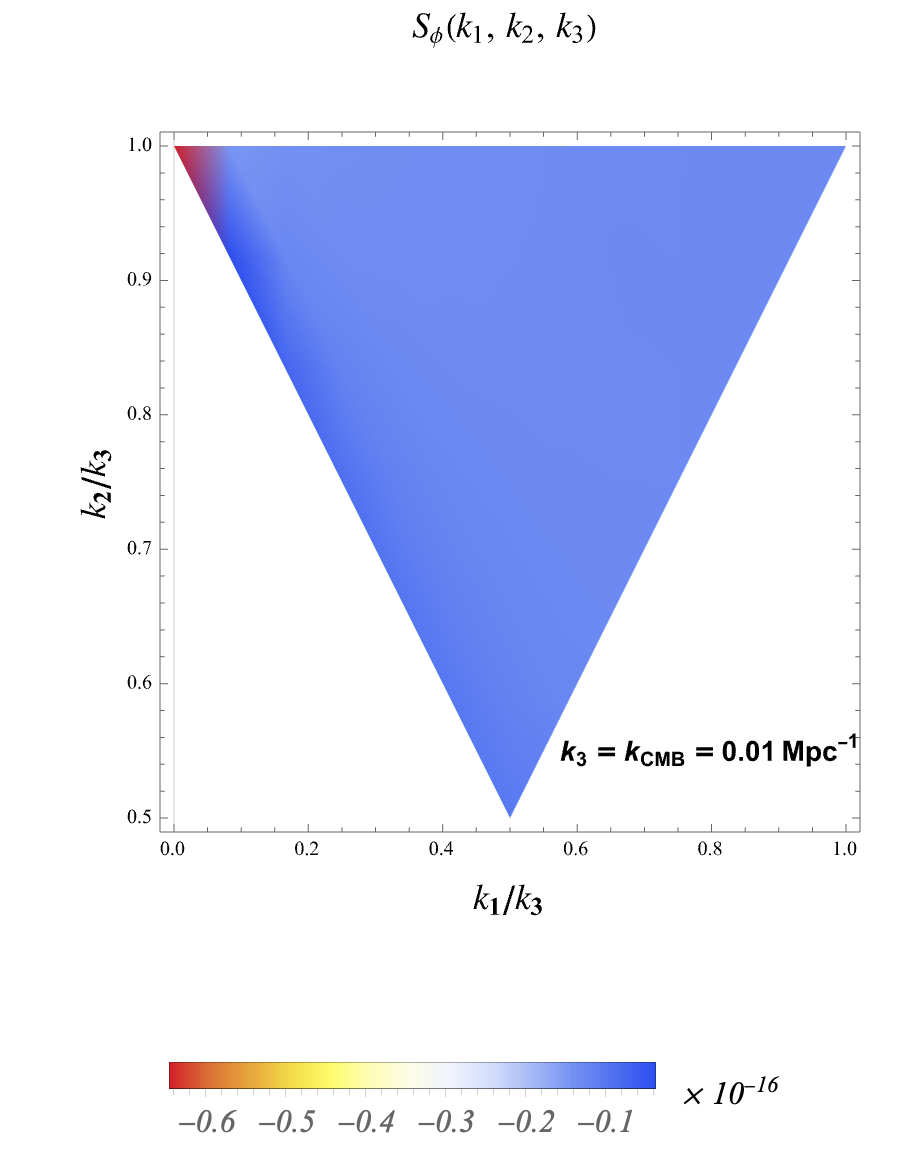}
\caption{
Dimensionless shape function $S_\phi(k_1,k_2,k_3)$
evaluated at fixed $k_3 = k_{\rm CMB} = 0.01\,\mathrm{Mpc}^{-1}$.
The horizontal and vertical axes correspond to the ratios
$k_1/k_3$ and $k_2/k_3$, restricted to the triangular domain
$k_1 \le k_2 \le k_3$.
The color scale indicates the amplitude of $S_\phi(k_1,k_2,k_3)$ in units of $10^{-16}$.
The signal is enhanced toward squeezed configurations
($k_1 \ll k_2 \simeq k_3$).
}
\label{fig:shape_cmb}
\end{figure}

The resulting shape function is shown in Fig.~\ref{fig:shape_cmb}. We find a clear enhancement toward squeezed configurations, $k_1 \ll k_2 \simeq k_3$, while the signal remains finite throughout the triangular domain. In standard single-field slow-roll inflation, the squeezed limit is constrained by Maldacena's consistency relation and is suppressed by the spectral tilt \cite{Maldacena:2002vr}. In the IWI scenario, however, scalar perturbations are not generated by a background inflaton field, but arise dynamically at second order from tensor fluctuations. For this reason, the standard single-field consistency relation does not directly apply, and the squeezed enhancement found here originates from the tensor-scalar quadratic coupling within the present setup.

This shape analysis identifies the configurations where the signal is largest, but it does not by itself quantify the observational relevance of the bispectrum. To address this point, we now turn to the overall amplitude of the signal and to the corresponding non-Gaussianity parameter.

\subsection{Amplitude and cutoff dependence of non-Gaussianity}

In this subsection, we explicitly derive the dependence of the non-Gaussianity parameter, $f_{\rm NL}$, on the ultraviolet cutoff, which we parametrize through $N_{\rm obs}$ as in Ref.~\cite{Traforetti:2025cax}.

We write the scalar bispectrum in the form
\begin{equation}
B_{\phi}(k_1,k_2,k_3;N_{\rm obs})
=
\frac{8^3}{k_1^2 k_2^2 k_3^2}
\left(\frac{H_{\rm inf}}{m_{\rm pl}}\right)^6
\,I_B(N_{\rm obs}) ,
\label{bispectrumnobs}
\end{equation}
where $I_B(N_{\rm obs})$ denotes the integral evaluated with the cutoff set by $N_{\rm obs}$.

Similarly, we write the dimensionless scalar power spectrum as
\begin{equation}
\Delta_{\phi}^2(k;N_{\rm obs})
=
\left(\frac{H_{\rm inf}}{m_{\rm pl}}\right)^4
\,I_P(N_{\rm obs}) ,
\label{deltaphi2nobs}
\end{equation}
with $I_P(N_{\rm obs})$ the corresponding integral.

Following Ref.~\cite{Traforetti:2025cax}, $H_{\rm inf}$ is determined by fixing the observed scalar amplitude at the pivot scale $k_{\rm CMB}$,
\begin{equation}
\Delta_{\phi}^2(k_{\rm CMB})=2.1\times 10^{-9},
\end{equation}
which gives the scaling relation
\begin{equation}
H_{\rm inf} =
m_{\rm pl}
\left(\frac{\Delta_{\phi}^2(k_{\rm CMB})}{I_P(N_{\rm obs})}\right)^{1/4}.
\label{hinfvsnobs}
\end{equation}
Substituting Eq.~\eqref{hinfvsnobs} into Eq.~\eqref{bispectrumnobs} yields the constrained amplitude
\begin{equation}
B_{\phi}(k_1,k_2,k_3;N_{\rm obs})
=
\frac{8^3}{k_1^2 k_2^2 k_3^2}
\left(\frac{\Delta_{\phi}^2(k_{\rm CMB})}{I_P(N_{\rm obs})}\right)^{3/2}
\,I_B(N_{\rm obs}) .
\label{bphinobs_final}
\end{equation}

Using the standard definition of the dimensional power spectrum,
\begin{equation}
P_{\phi}(k)=\frac{2\pi^2}{k^3}\,\Delta_{\phi}^2(k)\,,
\label{dimensionalPS}
\end{equation}
we define the non-Gaussianity parameter in the equilateral configuration, $k_1=k_2=k_3=k$, as $f_{\rm NL} \sim B_{\phi}/P_{\phi}^2$.
The explicit $k^{-6}$ factors in Eqs.~\eqref{bphinobs_final} and \eqref{dimensionalPS} then cancel, giving
\begin{equation}
f_{\rm NL}(N_{\rm obs}) =
\frac{2^7}{\pi^4}\,
\left[\Delta_{\phi}^2(k_{\rm CMB})\right]^{-1/2}\,
\frac{I_B(N_{\rm obs})}{\left[I_P(N_{\rm obs})\right]^{3/2}} \;.
\label{fnl_nobs}
\end{equation}
Equation~\eqref{fnl_nobs} shows explicitly that, within this parameterization, $f_{\rm NL}$ does not acquire any residual dependence on the observable external scale $k$, but depends only on the cutoff through $I_B$ and $I_P$.

\begin{figure}[h!]
    \centering
\includegraphics[width=\linewidth]{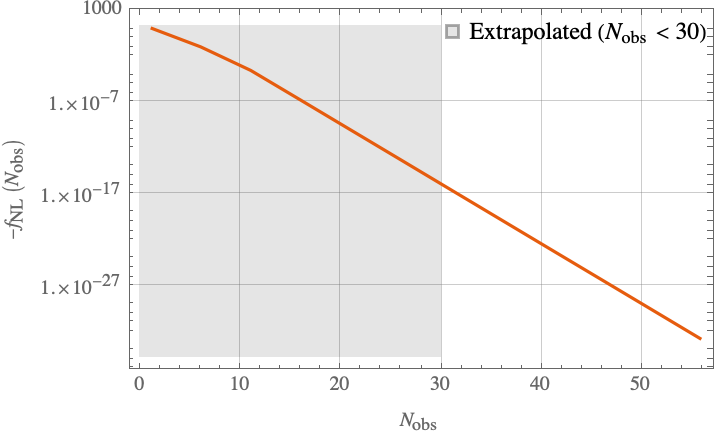}
\caption{
The negative value of the non-Gaussianity parameter, $f_{\rm NL}$, defined in Eq.~\eqref{fnl_nobs}, as a function of the cutoff parameter $N_{\rm obs}$. The shaded region, $N_{\rm obs}<30$, is displayed only as an extrapolation, whereas the physically relevant range is the non-shaded region. The figure shows that, once the scalar power spectrum is fixed to its observed amplitude, the bispectrum amplitude is rapidly suppressed as the cutoff is increased, following approximately an $e^{-3N_{\rm obs}/2}$ behaviour.
}
    \label{fig:runningNobs}
\end{figure}

We evaluate Eq.~\eqref{fnl_nobs} by scanning the parameter range $N_{\rm obs}\in[1,60]$. The extension to very small values of $N_{\rm obs}$ is shown only to illustrate the trend of the suppression outside the physically relevant regime, while the physically relevant range remains the one discussed in Ref.~\cite{Traforetti:2025cax} ($N_{\rm obs} \gtrsim 30$) \footnote{Recall that in Ref. \cite{Traforetti:2025cax} the regime $N_{\rm obs}\gtrsim30$ is the one which gives a tensor-to-scalar ratio consistent with current upper bounds at the CMB pivot scale \cite{Planck:2018jri}.} and is represented by the non-shaded region of Fig.~\ref{fig:runningNobs}. 

As shown in Fig.~\ref{fig:runningNobs}, the negative non-Gaussianity parameter $-f_{\rm NL}$ decreases rapidly as a function of $N_{\rm obs}$. Over the range considered here, the numerical result is fitted by an exponential suppression close to $e^{-3N_{\rm obs}/2}$. This behavior reflects the cumulative sourcing of scalar perturbations by a large number of tensor modes over many e-folds. As the number of contributing modes increases, the bispectrum becomes increasingly suppressed relative to the power spectrum, and the resulting scalar statistics become progressively closer to Gaussian.

This shows quantitatively that, once the scalar power spectrum is fixed to its observed amplitude, the induced bispectrum becomes increasingly suppressed as the ultraviolet cutoff is moved to larger values. The corresponding amplitude is extremely small throughout the physically relevant range. This is to be contrasted with current CMB constraints on primordial non-Gaussianity, which are of order unity to tens depending on the template \cite{Planck:2019kim}. In particular, Planck finds $f_{\rm NL}^{\rm local}=-0.9\pm5.1$ and $f_{\rm NL}^{\rm equil}=-26\pm47$ at $68\%$ confidence level \cite{Planck:2019kim}. By comparison, the values obtained here range from roughly $10^{-16}$ down to $10^{-36}$ in the physical range, rendering the IWI non-Gaussian signal observationally negligible.

%%%%%%%%%%%%%%%%%%%%%%%%%%%%%%%%%%%%%%%%%
\section{CONCLUSIONS}
\label{sec:conclusions}

In this work, we studied primordial non-Gaussianity in the framework of Inflation without an Inflaton (IWI), where scalar perturbations are generated dynamically at second order by tensor fluctuations in an exact de Sitter (dS) background. Extending the analysis of the scalar power spectrum presented in Ref.~\cite{Traforetti:2025cax}, we computed the scalar bispectrum induced by primordial gravitational waves.

We found that the bispectrum is logarithmically sensitive to the ultraviolet cutoff and that its shape function is enhanced toward squeezed configurations. At the same time, the overall amplitude of the signal at CMB scales is extremely small and therefore observationally negligible.
Although the generation mechanism is intrinsically nonlinear, the resulting non-Gaussian signal is strongly suppressed on observable scales. This suppression can be understood as a consequence of the continuous sourcing mechanism: scalar fluctuations are generated by the cumulative effect of many tensor modes over a large number of e-folds, so that the bispectrum becomes increasingly suppressed relative to the power spectrum and the resulting signal becomes progressively closer to Gaussian. Once the scalar power spectrum is fixed to its observed amplitude, the corresponding $f_{\rm NL}$ remains extremely small throughout the physically relevant range. In the parameterization adopted here, its amplitude depends on the ultraviolet cutoff through $N_{\rm obs}$, following the convention of Ref.~\cite{Traforetti:2025cax}. Scalar perturbations in the IWI scenario are therefore intrinsically non-Gaussian at the level of their generation, but effectively close to Gaussian at CMB scales.

%%%%%%%%%%%%%%%%%%%%%%%%%%%%%%%%%
\begin{acknowledgments}
MA acknowledges the hospitality of the Institute of Cosmos Sciences of the University of Barcelona (ICCUB), where this work was carried out. The project that gave rise to these results received the support of a fellowship from ``la Caixa'' Foundation (ID 100010434), awarded to MT with code LCF/BQ/DI24/12070012. The work of RJ is supported by the Simons Foundation. Funding for the work of MT and RJ was partially provided by project PID2022-141125NB-I00, and grant CEX2024-001451-M funded by MICIU/AEI/10.13039/501100011033. DB and SM acknowledge support from the COSMOS network (www.cosmosnet.it) through ASI (Italian Space Agency) Grants 2016-24-H.0, 2016-24-H.1-2018 and 2020-9-HH.0.
This work is supported in part by the MUR Departments of Excellence grant “Quantum Frontiers” of the Physics and
Astronomy Department of Padova, Italy.

\end{acknowledgments}

%%%%%%%%%%%%%%%%%%%%%%%%%%%%%%%%%%%%%%%%%
\bibliographystyle{apsrev4-2}
%\bibliography{references}

\begin{thebibliography}{26}%
\makeatletter
\providecommand \@ifxundefined [1]{%
 \@ifx{#1\undefined}
}%
\providecommand \@ifnum [1]{%
 \ifnum #1\expandafter \@firstoftwo
 \else \expandafter \@secondoftwo
 \fi
}%
\providecommand \@ifx [1]{%
 \ifx #1\expandafter \@firstoftwo
 \else \expandafter \@secondoftwo
 \fi
}%
\providecommand \natexlab [1]{#1}%
\providecommand \enquote  [1]{``#1''}%
\providecommand \bibnamefont  [1]{#1}%
\providecommand \bibfnamefont [1]{#1}%
\providecommand \citenamefont [1]{#1}%
\providecommand \href@noop [0]{\@secondoftwo}%
\providecommand \href [0]{\begingroup \@sanitize@url \@href}%
\providecommand \@href[1]{\@@startlink{#1}\@@href}%
\providecommand \@@href[1]{\endgroup#1\@@endlink}%
\providecommand \@sanitize@url [0]{\catcode `\\12\catcode `\$12\catcode
  `\&12\catcode `\#12\catcode `\^12\catcode `\_12\catcode `\%12\relax}%
\providecommand \@@startlink[1]{}%
\providecommand \@@endlink[0]{}%
\providecommand \url  [0]{\begingroup\@sanitize@url \@url }%
\providecommand \@url [1]{\endgroup\@href {#1}{\urlprefix }}%
\providecommand \urlprefix  [0]{URL }%
\providecommand \Eprint [0]{\href }%
\providecommand \doibase [0]{https://doi.org/}%
\providecommand \selectlanguage [0]{\@gobble}%
\providecommand \bibinfo  [0]{\@secondoftwo}%
\providecommand \bibfield  [0]{\@secondoftwo}%
\providecommand \translation [1]{[#1]}%
\providecommand \BibitemOpen [0]{}%
\providecommand \bibitemStop [0]{}%
\providecommand \bibitemNoStop [0]{.\EOS\space}%
\providecommand \EOS [0]{\spacefactor3000\relax}%
\providecommand \BibitemShut  [1]{\csname bibitem#1\endcsname}%
\let\auto@bib@innerbib\@empty
%</preamble>
\bibitem [{\citenamefont {Starobinsky}(1979)}]{Starobinsky:1979ty}%
  \BibitemOpen
  \bibfield  {author} {\bibinfo {author} {\bibfnamefont {A.~A.}\ \bibnamefont
  {Starobinsky}},\ }\href@noop {} {\bibfield  {journal} {\bibinfo  {journal}
  {JETP Lett.}\ }\textbf {\bibinfo {volume} {30}},\ \bibinfo {pages} {682}
  (\bibinfo {year} {1979})}\BibitemShut {NoStop}%
\bibitem [{\citenamefont {Starobinsky}(1980)}]{Starobinsky:1980te}%
  \BibitemOpen
  \bibfield  {author} {\bibinfo {author} {\bibfnamefont {A.~A.}\ \bibnamefont
  {Starobinsky}},\ }\href {https://doi.org/10.1016/0370-2693(80)90670-X}
  {\bibfield  {journal} {\bibinfo  {journal} {Phys. Lett. B}\ }\textbf
  {\bibinfo {volume} {91}},\ \bibinfo {pages} {99} (\bibinfo {year}
  {1980})}\BibitemShut {NoStop}%
\bibitem [{\citenamefont {Guth}(1981)}]{Guth:1980zm}%
  \BibitemOpen
  \bibfield  {author} {\bibinfo {author} {\bibfnamefont {A.~H.}\ \bibnamefont
  {Guth}},\ }\href {https://doi.org/10.1103/PhysRevD.23.347} {\bibfield
  {journal} {\bibinfo  {journal} {Phys. Rev. D}\ }\textbf {\bibinfo {volume}
  {23}},\ \bibinfo {pages} {347} (\bibinfo {year} {1981})}\BibitemShut
  {NoStop}%
\bibitem [{\citenamefont {Linde}(1982)}]{Linde:1981mu}%
  \BibitemOpen
  \bibfield  {author} {\bibinfo {author} {\bibfnamefont {A.~D.}\ \bibnamefont
  {Linde}},\ }\href {https://doi.org/10.1016/0370-2693(82)91219-9} {\bibfield
  {journal} {\bibinfo  {journal} {Phys. Lett. B}\ }\textbf {\bibinfo {volume}
  {108}},\ \bibinfo {pages} {389} (\bibinfo {year} {1982})}\BibitemShut
  {NoStop}%
\bibitem [{\citenamefont {Mukhanov}\ and\ \citenamefont
  {Chibisov}(1981)}]{Mukhanov:1981xt}%
  \BibitemOpen
  \bibfield  {author} {\bibinfo {author} {\bibfnamefont {V.~F.}\ \bibnamefont
  {Mukhanov}}\ and\ \bibinfo {author} {\bibfnamefont {G.~V.}\ \bibnamefont
  {Chibisov}},\ }\href@noop {} {\bibfield  {journal} {\bibinfo  {journal} {JETP
  Lett.}\ }\textbf {\bibinfo {volume} {33}},\ \bibinfo {pages} {532} (\bibinfo
  {year} {1981})}\BibitemShut {NoStop}%
\bibitem [{\citenamefont {Albrecht}\ and\ \citenamefont
  {Steinhardt}(1982)}]{Albrecht:1982wi}%
  \BibitemOpen
  \bibfield  {author} {\bibinfo {author} {\bibfnamefont {A.}~\bibnamefont
  {Albrecht}}\ and\ \bibinfo {author} {\bibfnamefont {P.~J.}\ \bibnamefont
  {Steinhardt}},\ }\href {https://doi.org/10.1103/PhysRevLett.48.1220}
  {\bibfield  {journal} {\bibinfo  {journal} {Phys. Rev. Lett.}\ }\textbf
  {\bibinfo {volume} {48}},\ \bibinfo {pages} {1220} (\bibinfo {year}
  {1982})}\BibitemShut {NoStop}%
\bibitem [{\citenamefont {Starobinsky}(1982)}]{Starobinsky:1982ee}%
  \BibitemOpen
  \bibfield  {author} {\bibinfo {author} {\bibfnamefont {A.~A.}\ \bibnamefont
  {Starobinsky}},\ }\href {https://doi.org/10.1016/0370-2693(82)90541-X}
  {\bibfield  {journal} {\bibinfo  {journal} {Phys. Lett. B}\ }\textbf
  {\bibinfo {volume} {117}},\ \bibinfo {pages} {175} (\bibinfo {year}
  {1982})}\BibitemShut {NoStop}%
\bibitem [{\citenamefont {Rubakov}\ \emph {et~al.}(1982)\citenamefont
  {Rubakov}, \citenamefont {Sazhin},\ and\ \citenamefont
  {Veryaskin}}]{Rubakov:1982df}%
  \BibitemOpen
  \bibfield  {author} {\bibinfo {author} {\bibfnamefont {V.~A.}\ \bibnamefont
  {Rubakov}}, \bibinfo {author} {\bibfnamefont {M.~V.}\ \bibnamefont
  {Sazhin}},\ and\ \bibinfo {author} {\bibfnamefont {A.~V.}\ \bibnamefont
  {Veryaskin}},\ }\href {https://doi.org/10.1016/0370-2693(82)90641-4}
  {\bibfield  {journal} {\bibinfo  {journal} {Phys. Lett. B}\ }\textbf
  {\bibinfo {volume} {115}},\ \bibinfo {pages} {189} (\bibinfo {year}
  {1982})}\BibitemShut {NoStop}%
\bibitem [{\citenamefont {Guth}\ and\ \citenamefont {Pi}(1982)}]{Guth:1982ec}%
  \BibitemOpen
  \bibfield  {author} {\bibinfo {author} {\bibfnamefont {A.~H.}\ \bibnamefont
  {Guth}}\ and\ \bibinfo {author} {\bibfnamefont {S.~Y.}\ \bibnamefont {Pi}},\
  }\href {https://doi.org/10.1103/PhysRevLett.49.1110} {\bibfield  {journal}
  {\bibinfo  {journal} {Phys. Rev. Lett.}\ }\textbf {\bibinfo {volume} {49}},\
  \bibinfo {pages} {1110} (\bibinfo {year} {1982})}\BibitemShut {NoStop}%
\bibitem [{\citenamefont {Linde}(1983)}]{Linde:1983gd}%
  \BibitemOpen
  \bibfield  {author} {\bibinfo {author} {\bibfnamefont {A.~D.}\ \bibnamefont
  {Linde}},\ }\href {https://doi.org/10.1016/0370-2693(83)90837-7} {\bibfield
  {journal} {\bibinfo  {journal} {Phys. Lett. B}\ }\textbf {\bibinfo {volume}
  {129}},\ \bibinfo {pages} {177} (\bibinfo {year} {1983})}\BibitemShut
  {NoStop}%
\bibitem [{\citenamefont {Kofman}\ \emph {et~al.}(1985)\citenamefont {Kofman},
  \citenamefont {Linde},\ and\ \citenamefont {Starobinsky}}]{Kofman:1985aw}%
  \BibitemOpen
  \bibfield  {author} {\bibinfo {author} {\bibfnamefont {L.~A.}\ \bibnamefont
  {Kofman}}, \bibinfo {author} {\bibfnamefont {A.~D.}\ \bibnamefont {Linde}},\
  and\ \bibinfo {author} {\bibfnamefont {A.~A.}\ \bibnamefont {Starobinsky}},\
  }\href {https://doi.org/10.1016/0370-2693(85)90381-8} {\bibfield  {journal}
  {\bibinfo  {journal} {Phys. Lett. B}\ }\textbf {\bibinfo {volume} {157}},\
  \bibinfo {pages} {361} (\bibinfo {year} {1985})}\BibitemShut {NoStop}%
\bibitem [{\citenamefont {Akrami}\ \emph
  {et~al.}(2020{\natexlab{a}})\citenamefont {Akrami} \emph
  {et~al.}}]{Planck:2018jri}%
  \BibitemOpen
  \bibfield  {author} {\bibinfo {author} {\bibfnamefont {Y.}~\bibnamefont
  {Akrami}} \emph {et~al.} (\bibinfo {collaboration} {Planck}),\ }\href
  {https://doi.org/10.1051/0004-6361/201833887} {\bibfield  {journal} {\bibinfo
   {journal} {Astron. Astrophys.}\ }\textbf {\bibinfo {volume} {641}},\
  \bibinfo {pages} {A10} (\bibinfo {year} {2020}{\natexlab{a}})},\ \Eprint
  {https://arxiv.org/abs/1807.06211} {arXiv:1807.06211 [astro-ph.CO]}
  \BibitemShut {NoStop}%
\bibitem [{\citenamefont {Aghanim}\ \emph {et~al.}(2020)\citenamefont {Aghanim}
  \emph {et~al.}}]{Planck:2018vyg}%
  \BibitemOpen
  \bibfield  {author} {\bibinfo {author} {\bibfnamefont {N.}~\bibnamefont
  {Aghanim}} \emph {et~al.} (\bibinfo {collaboration} {Planck}),\ }\href
  {https://doi.org/10.1051/0004-6361/201833910} {\bibfield  {journal} {\bibinfo
   {journal} {Astron. Astrophys.}\ }\textbf {\bibinfo {volume} {641}},\
  \bibinfo {pages} {A6} (\bibinfo {year} {2020})},\ \bibinfo {note} {[Erratum:
  Astron.Astrophys. 652, C4 (2021)]},\ \Eprint
  {https://arxiv.org/abs/1807.06209} {arXiv:1807.06209 [astro-ph.CO]}
  \BibitemShut {NoStop}%
\bibitem [{\citenamefont {Bertacca}\ \emph {et~al.}(2025)\citenamefont
  {Bertacca}, \citenamefont {Jimenez}, \citenamefont {Matarrese},\ and\
  \citenamefont {Ricciardone}}]{Bertacca:2024zfb}%
  \BibitemOpen
  \bibfield  {author} {\bibinfo {author} {\bibfnamefont {D.}~\bibnamefont
  {Bertacca}}, \bibinfo {author} {\bibfnamefont {R.}~\bibnamefont {Jimenez}},
  \bibinfo {author} {\bibfnamefont {S.}~\bibnamefont {Matarrese}},\ and\
  \bibinfo {author} {\bibfnamefont {A.}~\bibnamefont {Ricciardone}},\ }\href
  {https://doi.org/10.1103/vfny-pgc2} {\bibfield  {journal} {\bibinfo
  {journal} {Phys. Rev. Res.}\ }\textbf {\bibinfo {volume} {7}},\ \bibinfo
  {pages} {L032010} (\bibinfo {year} {2025})},\ \Eprint
  {https://arxiv.org/abs/2412.14265} {arXiv:2412.14265 [astro-ph.CO]}
  \BibitemShut {NoStop}%
\bibitem [{\citenamefont {Traforetti}\ \emph {et~al.}(2026)\citenamefont
  {Traforetti}, \citenamefont {Abdelaziz}, \citenamefont {Bertacca},
  \citenamefont {Jimenez}, \citenamefont {Matarrese},\ and\ \citenamefont
  {Ricciardone}}]{Traforetti:2025cax}%
  \BibitemOpen
  \bibfield  {author} {\bibinfo {author} {\bibfnamefont {M.}~\bibnamefont
  {Traforetti}}, \bibinfo {author} {\bibfnamefont {M.}~\bibnamefont
  {Abdelaziz}}, \bibinfo {author} {\bibfnamefont {D.}~\bibnamefont {Bertacca}},
  \bibinfo {author} {\bibfnamefont {R.}~\bibnamefont {Jimenez}}, \bibinfo
  {author} {\bibfnamefont {S.}~\bibnamefont {Matarrese}},\ and\ \bibinfo
  {author} {\bibfnamefont {A.}~\bibnamefont {Ricciardone}},\ }\href
  {https://doi.org/10.1103/cfg4-5p3s} {\bibfield  {journal} {\bibinfo
  {journal} {Phys. Rev. D}\ }\textbf {\bibinfo {volume} {113}},\ \bibinfo
  {pages} {023553} (\bibinfo {year} {2026})},\ \Eprint
  {https://arxiv.org/abs/2511.11808} {arXiv:2511.11808 [astro-ph.CO]}
  \BibitemShut {NoStop}%
\bibitem [{\citenamefont {Dvali}\ \emph {et~al.}(2017)\citenamefont {Dvali},
  \citenamefont {Gomez},\ and\ \citenamefont {Zell}}]{Dvali:2017eba}%
  \BibitemOpen
  \bibfield  {author} {\bibinfo {author} {\bibfnamefont {G.}~\bibnamefont
  {Dvali}}, \bibinfo {author} {\bibfnamefont {C.}~\bibnamefont {Gomez}},\ and\
  \bibinfo {author} {\bibfnamefont {S.}~\bibnamefont {Zell}},\ }\href
  {https://doi.org/10.1088/1475-7516/2017/06/028} {\bibfield  {journal}
  {\bibinfo  {journal} {JCAP}\ }\textbf {\bibinfo {volume} {06}},\ \bibinfo
  {pages} {028}},\ \Eprint {https://arxiv.org/abs/1701.08776} {arXiv:1701.08776
  [hep-th]} \BibitemShut {NoStop}%
\bibitem [{\citenamefont {Maldacena}(2003)}]{Maldacena:2002vr}%
  \BibitemOpen
  \bibfield  {author} {\bibinfo {author} {\bibfnamefont {J.~M.}\ \bibnamefont
  {Maldacena}},\ }\href {https://doi.org/10.1088/1126-6708/2003/05/013}
  {\bibfield  {journal} {\bibinfo  {journal} {JHEP}\ }\textbf {\bibinfo
  {volume} {05}},\ \bibinfo {pages} {013}},\ \Eprint
  {https://arxiv.org/abs/astro-ph/0210603} {arXiv:astro-ph/0210603}
  \BibitemShut {NoStop}%
\bibitem [{\citenamefont {Bartolo}\ \emph {et~al.}(2004)\citenamefont
  {Bartolo}, \citenamefont {Komatsu}, \citenamefont {Matarrese},\ and\
  \citenamefont {Riotto}}]{Bartolo:2004if}%
  \BibitemOpen
  \bibfield  {author} {\bibinfo {author} {\bibfnamefont {N.}~\bibnamefont
  {Bartolo}}, \bibinfo {author} {\bibfnamefont {E.}~\bibnamefont {Komatsu}},
  \bibinfo {author} {\bibfnamefont {S.}~\bibnamefont {Matarrese}},\ and\
  \bibinfo {author} {\bibfnamefont {A.}~\bibnamefont {Riotto}},\ }\href
  {https://doi.org/10.1016/j.physrep.2004.08.022} {\bibfield  {journal}
  {\bibinfo  {journal} {Phys. Rept.}\ }\textbf {\bibinfo {volume} {402}},\
  \bibinfo {pages} {103} (\bibinfo {year} {2004})},\ \Eprint
  {https://arxiv.org/abs/astro-ph/0406398} {arXiv:astro-ph/0406398}
  \BibitemShut {NoStop}%
\bibitem [{\citenamefont {Creminelli}\ and\ \citenamefont
  {Zaldarriaga}(2004)}]{Creminelli:2004yq}%
  \BibitemOpen
  \bibfield  {author} {\bibinfo {author} {\bibfnamefont {P.}~\bibnamefont
  {Creminelli}}\ and\ \bibinfo {author} {\bibfnamefont {M.}~\bibnamefont
  {Zaldarriaga}},\ }\href {https://doi.org/10.1088/1475-7516/2004/10/006}
  {\bibfield  {journal} {\bibinfo  {journal} {JCAP}\ }\textbf {\bibinfo
  {volume} {10}},\ \bibinfo {pages} {006}},\ \Eprint
  {https://arxiv.org/abs/astro-ph/0407059} {arXiv:astro-ph/0407059}
  \BibitemShut {NoStop}%
\bibitem [{\citenamefont {Bartolo}\ \emph {et~al.}(2002)\citenamefont
  {Bartolo}, \citenamefont {Matarrese},\ and\ \citenamefont
  {Riotto}}]{Bartolo:2001cw}%
  \BibitemOpen
  \bibfield  {author} {\bibinfo {author} {\bibfnamefont {N.}~\bibnamefont
  {Bartolo}}, \bibinfo {author} {\bibfnamefont {S.}~\bibnamefont {Matarrese}},\
  and\ \bibinfo {author} {\bibfnamefont {A.}~\bibnamefont {Riotto}},\ }\href
  {https://doi.org/10.1103/PhysRevD.65.103505} {\bibfield  {journal} {\bibinfo
  {journal} {Phys. Rev. D}\ }\textbf {\bibinfo {volume} {65}},\ \bibinfo
  {pages} {103505} (\bibinfo {year} {2002})},\ \Eprint
  {https://arxiv.org/abs/hep-ph/0112261} {arXiv:hep-ph/0112261} \BibitemShut
  {NoStop}%
\bibitem [{\citenamefont {Chen}\ \emph {et~al.}(2007)\citenamefont {Chen},
  \citenamefont {Huang}, \citenamefont {Kachru},\ and\ \citenamefont
  {Shiu}}]{Chen:2006nt}%
  \BibitemOpen
  \bibfield  {author} {\bibinfo {author} {\bibfnamefont {X.}~\bibnamefont
  {Chen}}, \bibinfo {author} {\bibfnamefont {M.-x.}\ \bibnamefont {Huang}},
  \bibinfo {author} {\bibfnamefont {S.}~\bibnamefont {Kachru}},\ and\ \bibinfo
  {author} {\bibfnamefont {G.}~\bibnamefont {Shiu}},\ }\href
  {https://doi.org/10.1088/1475-7516/2007/01/002} {\bibfield  {journal}
  {\bibinfo  {journal} {JCAP}\ }\textbf {\bibinfo {volume} {01}},\ \bibinfo
  {pages} {002}},\ \Eprint {https://arxiv.org/abs/hep-th/0605045}
  {arXiv:hep-th/0605045} \BibitemShut {NoStop}%
\bibitem [{\citenamefont {Akrami}\ \emph
  {et~al.}(2020{\natexlab{b}})\citenamefont {Akrami} \emph
  {et~al.}}]{Planck:2019kim}%
  \BibitemOpen
  \bibfield  {author} {\bibinfo {author} {\bibfnamefont {Y.}~\bibnamefont
  {Akrami}} \emph {et~al.} (\bibinfo {collaboration} {Planck}),\ }\href
  {https://doi.org/10.1051/0004-6361/201935891} {\bibfield  {journal} {\bibinfo
   {journal} {Astron. Astrophys.}\ }\textbf {\bibinfo {volume} {641}},\
  \bibinfo {pages} {A9} (\bibinfo {year} {2020}{\natexlab{b}})},\ \Eprint
  {https://arxiv.org/abs/1905.05697} {arXiv:1905.05697 [astro-ph.CO]}
  \BibitemShut {NoStop}%
\bibitem [{\citenamefont {Ade}\ \emph {et~al.}(2016)\citenamefont {Ade} \emph
  {et~al.}}]{Planck:2015zfm}%
  \BibitemOpen
  \bibfield  {author} {\bibinfo {author} {\bibfnamefont {P.~A.~R.}\
  \bibnamefont {Ade}} \emph {et~al.} (\bibinfo {collaboration} {Planck}),\
  }\href {https://doi.org/10.1051/0004-6361/201525836} {\bibfield  {journal}
  {\bibinfo  {journal} {Astron. Astrophys.}\ }\textbf {\bibinfo {volume}
  {594}},\ \bibinfo {pages} {A17} (\bibinfo {year} {2016})},\ \Eprint
  {https://arxiv.org/abs/1502.01592} {arXiv:1502.01592 [astro-ph.CO]}
  \BibitemShut {NoStop}%
\bibitem [{\citenamefont {Ade}\ \emph {et~al.}(2014)\citenamefont {Ade} \emph
  {et~al.}}]{Planck:2013wtn}%
  \BibitemOpen
  \bibfield  {author} {\bibinfo {author} {\bibfnamefont {P.~A.~R.}\
  \bibnamefont {Ade}} \emph {et~al.} (\bibinfo {collaboration} {Planck}),\
  }\href {https://doi.org/10.1051/0004-6361/201321554} {\bibfield  {journal}
  {\bibinfo  {journal} {Astron. Astrophys.}\ }\textbf {\bibinfo {volume}
  {571}},\ \bibinfo {pages} {A24} (\bibinfo {year} {2014})},\ \Eprint
  {https://arxiv.org/abs/1303.5084} {arXiv:1303.5084 [astro-ph.CO]}
  \BibitemShut {NoStop}%
\bibitem [{\citenamefont {Abdelaziz}\ \emph {et~al.}(2025)\citenamefont
  {Abdelaziz}, \citenamefont {Bari}, \citenamefont {Matarrese},\ and\
  \citenamefont {Ricciardone}}]{Abdelaziz:2025qpn}%
  \BibitemOpen
  \bibfield  {author} {\bibinfo {author} {\bibfnamefont {M.}~\bibnamefont
  {Abdelaziz}}, \bibinfo {author} {\bibfnamefont {P.}~\bibnamefont {Bari}},
  \bibinfo {author} {\bibfnamefont {S.}~\bibnamefont {Matarrese}},\ and\
  \bibinfo {author} {\bibfnamefont {A.}~\bibnamefont {Ricciardone}},\ }\href
  {https://doi.org/10.1103/bb22-pq2m} {\bibfield  {journal} {\bibinfo
  {journal} {Phys. Rev. D}\ }\textbf {\bibinfo {volume} {112}},\ \bibinfo
  {pages} {023505} (\bibinfo {year} {2025})},\ \Eprint
  {https://arxiv.org/abs/2504.07063} {arXiv:2504.07063 [astro-ph.CO]}
  \BibitemShut {NoStop}%
\bibitem [{\citenamefont {Espinosa}\ \emph {et~al.}(2018)\citenamefont
  {Espinosa}, \citenamefont {Racco},\ and\ \citenamefont
  {Riotto}}]{Espinosa:2018eve}%
  \BibitemOpen
  \bibfield  {author} {\bibinfo {author} {\bibfnamefont {J.~R.}\ \bibnamefont
  {Espinosa}}, \bibinfo {author} {\bibfnamefont {D.}~\bibnamefont {Racco}},\
  and\ \bibinfo {author} {\bibfnamefont {A.}~\bibnamefont {Riotto}},\ }\href
  {https://doi.org/10.1088/1475-7516/2018/09/012} {\bibfield  {journal}
  {\bibinfo  {journal} {JCAP}\ }\textbf {\bibinfo {volume} {09}},\ \bibinfo
  {pages} {012}},\ \Eprint {https://arxiv.org/abs/1804.07732} {arXiv:1804.07732
  [hep-ph]} \BibitemShut {NoStop}%
\end{thebibliography}
%%%%%%%%%%%%%%%%%%%%%%%%%%%%%%%%%%%%%%%%%


\appendix

\section{Bispectrum computation}
\label{appendixA}

In this Appendix, we compute the three-point function of scalar fluctuations,
$\langle \phi(\bm{k}_1)\phi(\bm{k}_2)\phi(\bm{k}_3)\rangle$, and we obtain the scalar bispectrum $B_\phi(k_1,k_2,k_3)$ from Eq.~\eqref{Threepnt}.

As in the main text, we consider the analytic solution
$\phi^{(2)} = {\cal F}_\chi/4$,
for which $\phi^{(2)}$ is constant on large scales, with ${\cal F}_\chi$ given in Eq.~\eqref{eq:F_chi}.
For simplicity, we retain only the first contribution in Eq.~\eqref{eq:F_chi}:
\begin{align}
\label{phi2}
\phi^{(2)} = \nabla^{-2}\!\Big[
&\frac{3}{4}\,\chi^{(1)\,lk,m}\chi^{(1)}_{kl,m}
+ \frac{1}{2}\,\chi^{(1)\,kl}\nabla^{2}\chi^{(1)}_{lk}
\nonumber\\
&\qquad
- \frac{1}{2}\,\chi^{(1)\,km}{}_{,l}\chi^{(1)\,l}{}_{m,k}
\Big].
\end{align}

The Fourier transform of a scalar field $\phi(\bm{x})$ is defined as
\begin{align}
\phi(\bm{x}) = \frac{1}{(2\pi)^3}
\int d^3\bm{k}\; \phi(\bm{k})\, e^{i \bm{k}\cdot \bm{x}} .
\end{align}
Similarly, the transverse--traceless tensor field $\chi^{(1)}_{ij}(\bm{x})$ can be written as
\begin{align}
\chi^{(1)}_{ij}(\bm{x}) = \frac{1}{(2\pi)^3}
\int d^3\bm{k}\;
\sum_{\lambda}
\chi^{\lambda}(\bm{k})\,
e^{\lambda}_{ij}(\bm{k})\,
e^{i \bm{k}\cdot \bm{x}} ,
\end{align}
where $\lambda$ labels the polarization states and
$e^{\lambda}_{ij}(\bm{k})$ are the polarization tensors.

Using these definitions, Eq.~\eqref{phi2} in Fourier space reads
\begin{align}
\label{fourierphi}
\phi^{(2)}(\bm{k}) &=
\frac{1}{k^2}
\sum_{s,t}
\int\!\frac{d^3\bm{p}}{(2\pi)^3}
\int\!\frac{d^3\bm{q}}{(2\pi)^3}\,
\delta^{(3)}\!\left(\bm{k}-(\bm{p}+\bm{q})\right)
\nonumber\\
&\quad\times
{\cal K}^{s,t}(\bm{p},\bm{q})\,
\chi^{s}(\bm{p})\,\chi^{t}(\bm{q}) ,
\end{align}
where the delta function enforces $\bm{q}=\bm{k}-\bm{p}$, while the kernel ${\cal K}^{s,t}(\bm{p},\bm{q})$ encoding the polarization structure is given by
\begin{align}
{\cal K}^{s,t}(\bm{p},\bm{q})
&=
\frac{3}{4}(\bm{p}\!\cdot\!\bm{q})\,
\epsilon^{s\,lk}(\hat{\bm{p}})\,\epsilon^{t}_{kl}(\hat{\bm{q}})
\nonumber\\
&\quad+
\frac{1}{4}(p^2+q^2)\,
\epsilon^{s\,kl}(\hat{\bm{p}})\,\epsilon^{t}_{lk}(\hat{\bm{q}})
\nonumber\\
&\quad-
\frac12\,p_l q_k\,
\epsilon^{s\,km}(\hat{\bm{p}})\,\epsilon^{t\,l}{}_{\!m}(\hat{\bm{q}})\;.
\label{eq: definition calK}
\end{align}
Using this, the three-point function of the scalar field can then be written as
\begin{widetext}
\begin{align}
\label{threepnt_app}
\bigl\langle \phi(\bm{k}_1)\phi(\bm{k}_2)\phi(\bm{k}_3)\bigr\rangle
&=
\frac{1}{8}\,
\frac{1}{k_1^2 k_2^2 k_3^2}
\sum_{s_i,t_i}
\int\!\prod_{i=1}^{3}\frac{d^3\bm{p}_i}{(2\pi)^3}
\left[
\prod_{i=1}^{3}
{\cal K}^{s_i,t_i}(\bm{p}_i,\bm{q}_i)
\right]
\bigl\langle
\chi^{s_1}(\bm{p}_1)\chi^{t_1}(\bm{q}_1)
\chi^{s_2}(\bm{p}_2)\chi^{t_2}(\bm{q}_2)
\chi^{s_3}(\bm{p}_3)\chi^{t_3}(\bm{q}_3)
\bigr\rangle \,,
\end{align}
\end{widetext}
where $\bm{q}_i=\bm{k}_i-\bm{p}_i$ for $i=1,2,3$.
In the expression above, the six-point function of the tensor modes is evaluated using Wick’s theorem.
Among the 15 possible pairings, those involving contractions within the same quadratic block generate disconnected contributions proportional to $\delta^{(3)}(\bm{k}_i)$ and do not contribute to the bispectrum.
The remaining eight contractions are connected and contribute to $B_\phi$.\\
The tensor two-point function defines the tensor power spectrum $\mathcal{P}_\chi$ as
\begin{align}
\langle \chi^{s_i}(\bm{p}_i)\chi^{t_j}(\bm{q}_j)\rangle
&=
(2\pi)^3\delta_{s_it_j}\delta^{(3)}(\bm{p}_i+\bm{q}_j)\mathcal{P}_\chi(p_i)
\nonumber\\
&=
(2\pi)^3\delta_{s_it_j}\delta^{(3)}(\bm{p}_i+\bm{q}_j)
\frac{2\pi^2}{p_i^3}\Delta^2_\chi(p_i)\,,
\end{align}
where $i\neq j$.\\
Due to permutation symmetry in $(\bm{k}_1,\bm{k}_2,\bm{k}_3)$ and the interchange
$\bm{p}_i \leftrightarrow \bm{q}_i$,
All eight connected Wick contractions give identical contributions after relabeling dummy variables.
It is therefore sufficient to evaluate one contraction and multiply the result by eight \cite{Espinosa:2018eve,Abdelaziz:2025qpn}.\\
We choose to evaluate the contraction $\langle 16\rangle\langle 23\rangle\langle 45\rangle$, which yields
\begin{widetext}
\begin{align}
\label{contraction}
\left < \chi^{s_1}(\bm{p}_1)\chi^{t_3}(\bm{q}_3) \right >
\left < \chi^{t_1}(\bm{q}_1)\chi^{s_2}(\bm{p}_2) \right >
\left < \chi^{t_2}(\bm{q}_2)\chi^{s_3}(\bm{p}_3) \right >
&=
(2\pi)^9\,
\delta_{s_1t_3}\delta_{t_1s_2}\delta_{t_2s_3}
\delta^{(3)}(\bm{k}_1+\bm{k}_2+\bm{k}_3)\,
\delta^{(3)}(\bm{p}_1+\bm{k}_3-\bm{p}_3)\, \nonumber\\
&\quad\times
\delta^{(3)}(\bm{k}_1-\bm{p}_1+\bm{p}_2)
(2\pi^2)^3
\prod_{i=1}^{3}\frac{\Delta_\chi^2(p_i)}{p_i^3}\; .
\end{align}
\end{widetext}
The Dirac deltas in Eq.~\eqref{contraction} fix the geometrical configuration of the six momenta.
Integrating over $d^3\bm{p}_2$ and $d^3\bm{p}_3$ leaves a single integral over $\bm{p}_1$, with
\begin{align}
\bm{p}_2=\bm{p}_1-\bm{k}_1, \qquad
\bm{p}_3=\bm{p}_1+\bm{k}_3 .
\label{eq: relation p1,p2,p3}
\end{align}
Therefore, the three-point function becomes
\begin{widetext}
\begin{align}
\label{finalthreepnt}
\bigl\langle \phi(\bm{k}_1)\phi(\bm{k}_2)\phi(\bm{k}_3)\bigr\rangle
&=
\frac{\pi^3}{k_1^2 k_2^2 k_3^2}
(2\pi)^3
\delta^{(3)}(\bm{k}_1+\bm{k}_2+\bm{k}_3)
\nonumber\\
&\quad\times
\int d^3\bm{p}_1\,
\frac{\Delta_\chi^2(p_1)}{p_1^3}
\frac{\Delta_\chi^2(p_2)}{p_2^3}
\frac{\Delta_\chi^2(p_3)}{p_3^3}
\left[\sum_{s_1,s_2,s_3}
{\cal K}^{s_1,s_2}(\bm{p}_1,-\bm{p}_2)
{\cal K}^{s_2,s_3}(\bm{p}_2,-\bm{p}_3)
{\cal K}^{s_3,s_1}(\bm{p}_3,-\bm{p}_1)\right]\,.
\end{align}
\end{widetext}
We define the kernel appearing in the square brackets as $K(\bm{p}_1;\bm{k}_1,\bm{k}_2,\bm{k}_3)$ and evaluate it explicitly in the following Appendix.

%%%%%%%%%%%%%%%%%%%%%%%%%%%%%%%%%%%%%%%%%
\section{Full kernel}
\label{sec:kernel}

In this Appendix, we compute explicitly the kernel $K(\bm{p}_1;\bm{k}_1,\bm{k}_2,\bm{k}_3)$ inside the square brackets in Eq.~\eqref{finalthreepnt}. Using the expression \eqref{eq: definition calK} for $\mathcal{K}^{s_i,t_i}(\bm{p_i},\bm{q_i})$, the kernel takes the following form
\begin{widetext}
\begin{eqnarray}
K(\bm{p}_1;\bm{k}_1,\bm{k}_2,\bm{k}_3)&=& \sum_{s_1,s_2,s_3}\left[-
\frac{3}{4}\,\bm{p_1}\cdot \bm{p_2} \,\epsilon^{s_1 lk}(\hat{\bm{p}}_{\bm 1})\,\epsilon^{s_2}_{kl}(-\hat{\bm{p}}_2)
\;+\;\frac{1}{4}\,(p_1^2+p_2^2)\,\epsilon^{s_1 kl}(\hat{\bm{p}}_{\bm 1})\,\epsilon^{s_2}_{lk}(-\hat{\bm{p}}_2)+\;\frac{1}{2}\,p_{1l}\,p_{2k}\,\epsilon^{s_1 km}(\hat{\bm{p}}_{\bm 1})\,\epsilon^{s_2 l}_{\; \; \; \; m}(-\hat{\bm{p}}_2)
\right]\nonumber\\
&&\times\left[-\frac{3}{4}\,\bm{p_2}\cdot\bm{p_3}\,\epsilon^{s_2 uv}(\hat{\bm{p}}_{\bm 2})\,\epsilon^{s_3}_{vu}(-\hat{\bm{p}}_3)+\;\frac{1}{4}\,(p_2^2 +p_3^2)\,\epsilon^{s_2 vu}(\hat{\bm{p}}_{\bm 2})\,\epsilon^{s_3}_{uv}(-\hat{\bm{p}}_3)
+\;\frac{1}{2}\,p_{2u}\,p_{3v}\,\epsilon^{s_2 vr}(\hat{\bm{p}}_{\bm 2})\,\epsilon^{s_3 u}_{\; \; \; \; \; r}(-\hat{\bm{p}}_3)\right]\nonumber\\
&&\times\left[-\frac{3}{4}\,\bm{p_3}\cdot\bm{p_1}\,\epsilon^{s_3 ij}(\hat{\bm{p}}_{\bm 3})\,\epsilon^{s_1}_{ji}(-\hat{\bm{p}}_1)+\;\frac{1}{4}\,(p_3^2+p_1^2)\,\epsilon^{s_3 ji}(\hat{\bm{p}}_{\bm 3})\,\epsilon^{s_1}_{ij}(-\hat{\bm{p}}_1)+\;\frac{1}{2}\,p_{3i}\,p_{1j}\,\epsilon^{s_3 jz}(\hat{\bm{p}}_{\bm 3})\,\epsilon^{s_1 i}_{\; \; \; \; \;z}(-\hat{\bm{p}}_1)\right]\,.
\label{eq: explicit kernel K}
\end{eqnarray}
\end{widetext}

To elaborate Eq.~\eqref{eq: explicit kernel K}, we need an explicit expression for the polarization tensors $\epsilon^\lambda_{ab}(\hat{\bm{k}})$, with $\lambda$ being the polarization index.

We define the ``plus'' $(+)$ and ``cross'' $(\times)$ polarization tensors as
\begin{equation}
    \epsilon_{a b}^{+}(\hat{\bm{k}})=\frac{\hat{\boldsymbol{u}}_{a} \hat{\boldsymbol{u}}_{b}-\hat{\boldsymbol{v}}_{a} \hat{\boldsymbol{v}}_{b}}{\sqrt{2}}\,, \quad \epsilon_{a b}^{\times}(\hat{\bm{k}})=\frac{\hat{\boldsymbol{u}}_{a} \hat{\boldsymbol{v}}_{b}+\hat{\boldsymbol{v}}_{a} \hat{\boldsymbol{u}}_{b}}{\sqrt{2}}  \,,
\end{equation}
where $\boldsymbol{u}$ and $\boldsymbol{v}$ are vectors orthogonal to the direction of propagation of the tensor modes, i.e., $\bm{k}$. More precisely, $\{\bm{k},\boldsymbol{u},\boldsymbol{v}\}$ form an orthonormal basis.
The polarization tensors obey
\begin{equation}
\begin{array}{ccc}
\epsilon_{a b}^{\lambda}(\boldsymbol{\hat{k}}) \epsilon^{a b}_{\lambda'}(\boldsymbol{\hat{k}})=\delta^{\lambda}_{\lambda'}\,, & \epsilon_{a b}^{\lambda}(\boldsymbol{\hat{k}})\delta^{ab}=0\,, \\ k^a\epsilon_{a b}^{\lambda}(\boldsymbol{\hat{k}})=0\,,
 & \epsilon_{a b}^{\lambda}(\boldsymbol{\hat{k}})=\epsilon_{a b}^{\lambda*}(\boldsymbol{\hat{k}})=\epsilon_{a b}^{\lambda}(-\boldsymbol{\hat{k}})\,.\\
\label{eq: properties polarization tensors}
\end{array}
\end{equation}
Using the previous properties, it can be shown that the polarization tensors satisfy the following identity
\begin{align} \label{polid}
2 \sum_{\lambda} \epsilon^\lambda_{i j}(\hat{\bm{k}}) \epsilon_{a b}^{\lambda\,*}(\hat{\bm{k}})=& \left(\delta_{i a}-\hat{\bm{k}}_{i} \hat{\bm{k}}_{a}\right)\left(\delta_{j b}-\hat{\bm{k}}_{j} \hat{\bm{k}}_{b}\right)
\nonumber \\ &+\left(\delta_{i b}-\hat{\bm{k}}_{i} \hat{\bm{k}}_{b}\right)\left(\delta_{j a}-\hat{\bm{k}}_{j} \hat{\bm{k}}_{a}\right)
\nonumber \\&-\left(\delta_{i j}-\hat{\bm{k}}_{i} \hat{\bm{k}}_{j}\right)\left(\delta_{a b}-\hat{\bm{k}}_{a} \hat{\bm{k}}_{b}\right) \,.
\end{align}
Therefore, using the fact that $\epsilon_{a b}^{\lambda}(-\hat{\bm{k}})=\epsilon_{a b}^{\lambda\,*}(\hat{\bm{k}})$, we can insert the identity \eqref{polid} to write explicitly the contractions of the polarization tensors appearing in the full kernel \eqref{eq: explicit kernel K}. After some tedious calculations, we arrive at the following final form for $K(\bm{p}_1;\bm{k}_1,\bm{k}_2,\bm{k}_3)$:
\begin{widetext}
\begin{align}
\label{eq: kernel K explicitly}
K(\bm{p}_1;\bm{k}_1,\bm{k}_2,\bm{k}_3) =& \dfrac{1}{8} \Biggl\{ \mathcal{I}_1(\bm{p}_1,\bm{p}_2,\bm{p}_3) \biggl[
-\dfrac{27}{64} (\bm{p}_1 \cdot \bm{p}_2)(\bm{p}_2 \cdot \bm{p}_3)(\bm{p}_3 \cdot \bm{p}_1)
+ \dfrac{9}{64} (\bm{p}_1 \cdot \bm{p}_2)(\bm{p}_2 \cdot \bm{p}_3)(p_1^2 + p_3^2) \nonumber\\
&\qquad\qquad\qquad\qquad + \dfrac{9}{64} (\bm{p}_1 \cdot \bm{p}_2)(\bm{p}_3 \cdot \bm{p}_1)(p_2^2 + p_3^2)
+ \dfrac{9}{64} (\bm{p}_2 \cdot \bm{p}_3)(\bm{p}_3 \cdot \bm{p}_1)(p_1^2 + p_2^2) \nonumber\\
&\qquad\qquad\qquad\qquad - \dfrac{3}{64} (\bm{p}_1 \cdot \bm{p}_2)(p_2^2 + p_3^2)(p_3^2 + p_1^2)
- \dfrac{3}{64} (\bm{p}_2 \cdot \bm{p}_3)(p_1^2 + p_2^2)(p_3^2 + p_1^2) \nonumber\\
&\qquad\qquad\qquad\qquad - \dfrac{3}{64} (\bm{p}_3 \cdot \bm{p}_1)(p_1^2 + p_2^2)(p_2^2 + p_3^2)
+ \dfrac{1}{64} (p_1^2 + p_2^2)(p_1^2 + p_3^2)(p_2^2 + p_3^2) \biggr]\nonumber \\
& \quad+ \mathcal{I}_2(\bm{p}_1,\bm{p}_2,\bm{p}_3) \biggl[ \dfrac{1}{32} \, p_{1} p_{3} \Bigl(
9 (\bm{p}_1 \cdot \bm{p}_2)(\bm{p}_2 \cdot \bm{p}_3)
- 3 (\bm{p}_1 \cdot \bm{p}_2)(p_2^2 + p_3^2)- 3 (\bm{p}_2 \cdot \bm{p}_3)(p_1^2 + p_2^2) \nonumber\\
&\qquad\qquad\qquad\qquad\quad + (p_1^2 + p_2^2)(p_2^2 + p_3^2)
\Bigr) \biggr] \nonumber\\
&\quad + \mathcal{I}_3(\bm{p}_1,\bm{p}_2,\bm{p}_3) \biggl[ \dfrac{1}{32} \, p_{2} p_{3} \Bigl(
9 (\bm{p}_1 \cdot \bm{p}_2)(\bm{p}_3 \cdot \bm{p}_1)
- 3 (\bm{p}_1 \cdot \bm{p}_2)(p_1^2 + p_3^2)- 3 (\bm{p}_3 \cdot \bm{p}_1)(p_1^2 + p_2^2) \nonumber\\
&\qquad\qquad\qquad\qquad\quad  + (p_1^2 + p_2^2)(p_1^2 + p_3^2)\Bigr) \biggr] \nonumber\\
& \quad + \mathcal{I}_4(\bm{p}_1,\bm{p}_2,\bm{p}_3) \biggl[ \dfrac{1}{16} \, p_{1} p_{2} p_{3}^2
\Bigl(p_1^2 - 3 (\bm{p}_1 \cdot \bm{p}_2) + p_2^2 \Bigr) \biggr] \nonumber\\
&\quad + \mathcal{I}_5(\bm{p}_1,\bm{p}_2,\bm{p}_3) \biggl[ \dfrac{1}{32} \, p_{1} p_{2} \Bigl(
9 (\bm{p}_2 \cdot \bm{p}_3)(\bm{p}_3 \cdot \bm{p}_1)
- 3 (\bm{p}_2 \cdot \bm{p}_3)(p_1^2 + p_3^2)- 3 (\bm{p}_3 \cdot \bm{p}_1)(p_2^2 + p_3^2)\nonumber \\
&\qquad\qquad\qquad\qquad\quad + (p_1^2 + p_3^2)(p_2^2 + p_3^2)
\Bigr) \biggr] \nonumber\\
&\quad + \mathcal{I}_6(\bm{p}_1,\bm{p}_2,\bm{p}_3) \biggl[ \dfrac{1}{16} \, p_{1}^2  p_{2} p_{3}
\Bigl(p_2^2 - 3 (\bm{p}_2 \cdot \bm{p}_3) + p_3^2 \Bigr) \biggr] \nonumber\\
&\quad + \mathcal{I}_7(\bm{p}_1,\bm{p}_2,\bm{p}_3) \biggl[ \dfrac{1}{16} \, p_{1} p_{2}^2  p_{3}
\Bigl(p_1^2 + p_3^2 - 3 (\bm{p}_3 \cdot \bm{p}_1) \Bigr) \biggr]
+ \mathcal{I}_8(\bm{p}_1,\bm{p}_2,\bm{p}_3)\biggl[ \dfrac{1}{8} p_1^2 p_2^2 p_3^2 \biggr] \Biggr\} .
\end{align}
\end{widetext}
where recall that the internal momenta $\bm{p}_i$ are related as in \eqref{eq: relation p1,p2,p3}.
In \eqref{eq: kernel K explicitly}, we collected the similar terms $\mathcal{I}_\alpha(\bm{p}_1,\bm{p}_2,\bm{p}_3)$ with $\alpha=1-8$, for simplicity.
The expressions for the different contractions $\mathcal{I}_\alpha(\bm{p}_1,\bm{p}_2,\bm{p}_3)$ appearing in Eq.~\eqref{eq: kernel K explicitly} are provided below. To simplify the output, we already used the fact that the $\hat{\bm p}_i$ are unit vectors, so $\hat{\bm p}_i^2=1$.

\begin{widetext}
\begin{align}
\mathcal{I}_1(\bm{p}_1,\bm{p}_2,\bm{p}_3) &= -4(\hat{\bm p}_1\!\cdot\!\hat{\bm p}_2)^3(\hat{\bm p}_1\!\cdot\!\hat{\bm p}_3)(\hat{\bm p}_2\!\cdot\!\hat{\bm p}_3) 
+ 5(\hat{\bm p}_1\!\cdot\!\hat{\bm p}_2)^2(\hat{\bm p}_1\!\cdot\!\hat{\bm p}_3)^2 
+ (\hat{\bm p}_1\!\cdot\!\hat{\bm p}_2)^2(\hat{\bm p}_1\!\cdot\!\hat{\bm p}_3)^2(\hat{\bm p}_2\!\cdot\!\hat{\bm p}_3)^2 + 5(\hat{\bm p}_1\!\cdot\!\hat{\bm p}_2)^2(\hat{\bm p}_2\!\cdot\!\hat{\bm p}_3)^2
 \nonumber \\
&\quad - 4(\hat{\bm p}_1\!\cdot\!\hat{\bm p}_2)(\hat{\bm p}_1\!\cdot\!\hat{\bm p}_3)(\hat{\bm p}_2\!\cdot\!\hat{\bm p}_3)^3 - 4(\hat{\bm p}_1\!\cdot\!\hat{\bm p}_2)(\hat{\bm p}_1\!\cdot\!\hat{\bm p}_3)^3(\hat{\bm p}_2\!\cdot\!\hat{\bm p}_3)+ 4(\hat{\bm p}_1\!\cdot\!\hat{\bm p}_2)(\hat{\bm p}_1\!\cdot\!\hat{\bm p}_3)(\hat{\bm p}_2\!\cdot\!\hat{\bm p}_3)
+ 5(\hat{\bm p}_1\!\cdot\!\hat{\bm p}_3)^2(\hat{\bm p}_2\!\cdot\!\hat{\bm p}_3)^2
\nonumber\\
&\quad+ 2(\hat{\bm p}_1\!\cdot\!\hat{\bm p}_2)^4 + (\hat{\bm p}_1\!\cdot\!\hat{\bm p}_2)^2 + 2(\hat{\bm p}_1\!\cdot\!\hat{\bm p}_3)^4 + (\hat{\bm p}_1\!\cdot\!\hat{\bm p}_3)^2
+ 2(\hat{\bm p}_2\!\cdot\!\hat{\bm p}_3)^4 + (\hat{\bm p}_2\!\cdot\!\hat{\bm p}_3)^2 - 1\,.
\end{align}
\begin{align}
\mathcal{I}_2(\bm{p}_1,\bm{p}_2,\bm{p}_3) &= -4(\hat{\bm p}_1\!\cdot\!\hat{\bm p}_2)(\hat{\bm p}_2\!\cdot\!\hat{\bm p}_3)(\hat{\bm p}_1\!\cdot\!\hat{\bm p}_3)^4
+ (\hat{\bm p}_1\!\cdot\!\hat{\bm p}_3)^3\Bigl[(\hat{\bm p}_1\!\cdot\!\hat{\bm p}_2)^2 \bigl((\hat{\bm p}_2\!\cdot\!\hat{\bm p}_3)^2+5\bigr)
+ 5(\hat{\bm p}_2\!\cdot\!\hat{\bm p}_3)^2-3\Bigr] \nonumber \\ &\quad -4(\hat{\bm p}_1\!\cdot\!\hat{\bm p}_2)(\hat{\bm p}_2\!\cdot\!\hat{\bm p}_3)(\hat{\bm p}_1\!\cdot\!\hat{\bm p}_3)^2
\Bigl[(\hat{\bm p}_1\!\cdot\!\hat{\bm p}_2)^2+(\hat{\bm p}_2\!\cdot\!\hat{\bm p}_3)^2-1\Bigr] -4(\hat{\bm p}_1\!\cdot\!\hat{\bm p}_2)(\hat{\bm p}_2\!\cdot\!\hat{\bm p}_3)
\bigl[(\hat{\bm p}_1\!\cdot\!\hat{\bm p}_2)^2+(\hat{\bm p}_2\!\cdot\!\hat{\bm p}_3)^2\bigr] \nonumber\\
&\quad+ 2(\hat{\bm p}_1\!\cdot\!\hat{\bm p}_3)^5 + (\hat{\bm p}_1\!\cdot\!\hat{\bm p}_3)\Bigl[(\hat{\bm p}_1\!\cdot\!\hat{\bm p}_2)^2
\bigl(11(\hat{\bm p}_2\!\cdot\!\hat{\bm p}_3)^2-5\bigr) + 2(\hat{\bm p}_1\!\cdot\!\hat{\bm p}_2)^4  + 2(\hat{\bm p}_2\!\cdot\!\hat{\bm p}_3)^4 - 5(\hat{\bm p}_2\!\cdot\!\hat{\bm p}_3)^2 + 1\Bigr]\,.
\end{align}
\begin{align}
\mathcal{I}_3(\bm{p}_1,\bm{p}_2,\bm{p}_3) &= 2 (\hat{\bm p}_1\!\cdot\!\hat{\bm p}_2)^4 (\hat{\bm p}_2\!\cdot\!\hat{\bm p}_3)-4 (\hat{\bm p}_1\!\cdot\!\hat{\bm p}_2)^3 (\hat{\bm p}_1\!\cdot\!\hat{\bm p}_3)\bigl((\hat{\bm p}_2\!\cdot\!\hat{\bm p}_3)^2+1\bigr)
+ (\hat{\bm p}_1\!\cdot\!\hat{\bm p}_2)^2 (\hat{\bm p}_2\!\cdot\!\hat{\bm p}_3)\Bigl[(\hat{\bm p}_1\!\cdot\!\hat{\bm p}_3)^2
\bigl((\hat{\bm p}_2\!\cdot\!\hat{\bm p}_3)^2+11\bigr)  \nonumber \\
&\quad + 5\bigl((\hat{\bm p}_2\!\cdot\!\hat{\bm p}_3)^2-1\bigr)\Bigr]-4 (\hat{\bm p}_1\!\cdot\!\hat{\bm p}_2)\Bigl[(\hat{\bm p}_1\!\cdot\!\hat{\bm p}_3)^3\bigl((\hat{\bm p}_2\!\cdot\!\hat{\bm p}_3)^2+1\bigr)
+ (\hat{\bm p}_1\!\cdot\!\hat{\bm p}_3)\,(\hat{\bm p}_2\!\cdot\!\hat{\bm p}_3)^2\bigl((\hat{\bm p}_2\!\cdot\!\hat{\bm p}_3)^2-1\bigr)\Bigr] \nonumber\\
&\quad+ (\hat{\bm p}_2\!\cdot\!\hat{\bm p}_3)\Bigl[5(\hat{\bm p}_1\!\cdot\!\hat{\bm p}_3)^2\bigl((\hat{\bm p}_2\!\cdot\!\hat{\bm p}_3)^2-1\bigr)
+ 2(\hat{\bm p}_1\!\cdot\!\hat{\bm p}_3)^4 + 2(\hat{\bm p}_2\!\cdot\!\hat{\bm p}_3)^4 - 3(\hat{\bm p}_2\!\cdot\!\hat{\bm p}_3)^2 + 1\Bigr]\,.
\end{align}
\begin{align}
\mathcal{I}_4(\bm{p}_1,\bm{p}_2,\bm{p}_3)&= 2 (\hat{\bm p}_1\!\cdot\!\hat{\bm p}_2)^4 (\hat{\bm p}_1\!\cdot\!\hat{\bm p}_3)\,(\hat{\bm p}_2\!\cdot\!\hat{\bm p}_3)
-4 (\hat{\bm p}_1\!\cdot\!\hat{\bm p}_2)^3 \Bigl[(\hat{\bm p}_1\!\cdot\!\hat{\bm p}_3)^2\bigl((\hat{\bm p}_2\!\cdot\!\hat{\bm p}_3)^2+1\bigr)
+ (\hat{\bm p}_2\!\cdot\!\hat{\bm p}_3)^2-1\Bigr] \nonumber \\
&\quad + (\hat{\bm p}_1\!\cdot\!\hat{\bm p}_2)^2 (\hat{\bm p}_1\!\cdot\!\hat{\bm p}_3)\,(\hat{\bm p}_2\!\cdot\!\hat{\bm p}_3)
\Bigl[(\hat{\bm p}_1\!\cdot\!\hat{\bm p}_3)^2\bigl((\hat{\bm p}_2\!\cdot\!\hat{\bm p}_3)^2+11\bigr) 
+ 11\bigl((\hat{\bm p}_2\!\cdot\!\hat{\bm p}_3)^2-1\bigr)\Bigr]\nonumber \\
&\quad -2 (\hat{\bm p}_1\!\cdot\!\hat{\bm p}_2)\Bigl[2(\hat{\bm p}_1\!\cdot\!\hat{\bm p}_3)^4\bigl((\hat{\bm p}_2\!\cdot\!\hat{\bm p}_3)^2+1\bigr)
+ (\hat{\bm p}_1\!\cdot\!\hat{\bm p}_3)^2\bigl(2(\hat{\bm p}_2\!\cdot\!\hat{\bm p}_3)^4+(\hat{\bm p}_2\!\cdot\!\hat{\bm p}_3)^2-3\bigr) + 2(\hat{\bm p}_2\!\cdot\!\hat{\bm p}_3)^4-3(\hat{\bm p}_2\!\cdot\!\hat{\bm p}_3)^2+1\Bigr]  \nonumber\\
&\quad+ (\hat{\bm p}_1\!\cdot\!\hat{\bm p}_3)\,(\hat{\bm p}_2\!\cdot\!\hat{\bm p}_3)\Bigl[5(\hat{\bm p}_1\!\cdot\!\hat{\bm p}_3)^2
\bigl((\hat{\bm p}_2\!\cdot\!\hat{\bm p}_3)^2-1\bigr) + 2(\hat{\bm p}_1\!\cdot\!\hat{\bm p}_3)^4 + 2(\hat{\bm p}_2\!\cdot\!\hat{\bm p}_3)^4 - 5(\hat{\bm p}_2\!\cdot\!\hat{\bm p}_3)^2 + 3\Bigr]\,.
\end{align}

\begin{align}
\mathcal{I}_5(\bm{p}_1,\bm{p}_2,\bm{p}_3)&= 2(\hat{\bm p}_1\!\cdot\!\hat{\bm p}_2)^5
-4(\hat{\bm p}_1\!\cdot\!\hat{\bm p}_2)^4(\hat{\bm p}_1\!\cdot\!\hat{\bm p}_3)(\hat{\bm p}_2\!\cdot\!\hat{\bm p}_3)
+ (\hat{\bm p}_1\!\cdot\!\hat{\bm p}_2)^3\Bigl[(\hat{\bm p}_1\!\cdot\!\hat{\bm p}_3)^2\bigl((\hat{\bm p}_2\!\cdot\!\hat{\bm p}_3)^2+5\bigr)
+ 5(\hat{\bm p}_2\!\cdot\!\hat{\bm p}_3)^2-3\Bigr]
 \nonumber \\
&\quad-4(\hat{\bm p}_1\!\cdot\!\hat{\bm p}_3)(\hat{\bm p}_2\!\cdot\!\hat{\bm p}_3)\bigl[(\hat{\bm p}_1\!\cdot\!\hat{\bm p}_3)^2+(\hat{\bm p}_2\!\cdot\!\hat{\bm p}_3)^2\bigr] -4(\hat{\bm p}_1\!\cdot\!\hat{\bm p}_2)^2(\hat{\bm p}_1\!\cdot\!\hat{\bm p}_3)(\hat{\bm p}_2\!\cdot\!\hat{\bm p}_3)
\bigl[(\hat{\bm p}_1\!\cdot\!\hat{\bm p}_3)^2+(\hat{\bm p}_2\!\cdot\!\hat{\bm p}_3)^2-1\bigr]
 \nonumber\\
&\quad+ (\hat{\bm p}_1\!\cdot\!\hat{\bm p}_2)\Bigl[(\hat{\bm p}_1\!\cdot\!\hat{\bm p}_3)^2\bigl(11(\hat{\bm p}_2\!\cdot\!\hat{\bm p}_3)^2-5\bigr)
+ 2(\hat{\bm p}_1\!\cdot\!\hat{\bm p}_3)^4 + 2(\hat{\bm p}_2\!\cdot\!\hat{\bm p}_3)^4  - 5(\hat{\bm p}_2\!\cdot\!\hat{\bm p}_3)^2 + 1\Bigr]\,.
\end{align}

\begin{align}
\mathcal{I}_6(\bm{p}_1,\bm{p}_2,\bm{p}_3) &= 2(\hat{\bm p}_1\!\cdot\!\hat{\bm p}_2)^5(\hat{\bm p}_1\!\cdot\!\hat{\bm p}_3)
-4(\hat{\bm p}_1\!\cdot\!\hat{\bm p}_2)^4\bigl((\hat{\bm p}_1\!\cdot\!\hat{\bm p}_3)^2+1\bigr)(\hat{\bm p}_2\!\cdot\!\hat{\bm p}_3)
+ (\hat{\bm p}_1\!\cdot\!\hat{\bm p}_2)^3(\hat{\bm p}_1\!\cdot\!\hat{\bm p}_3)\Bigl[(\hat{\bm p}_1\!\cdot\!\hat{\bm p}_3)^2
\bigl((\hat{\bm p}_2\!\cdot\!\hat{\bm p}_3)^2+5\bigr)
 \nonumber \\
&\quad + 11(\hat{\bm p}_2\!\cdot\!\hat{\bm p}_3)^2-5\Bigr] -2(\hat{\bm p}_1\!\cdot\!\hat{\bm p}_2)^2(\hat{\bm p}_2\!\cdot\!\hat{\bm p}_3)\Bigl[(\hat{\bm p}_1\!\cdot\!\hat{\bm p}_3)^2
\bigl(2(\hat{\bm p}_2\!\cdot\!\hat{\bm p}_3)^2+1\bigr) + 2(\hat{\bm p}_1\!\cdot\!\hat{\bm p}_3)^4 + 2(\hat{\bm p}_2\!\cdot\!\hat{\bm p}_3)^2-3\Bigr]
 \nonumber \\
&\quad + (\hat{\bm p}_1\!\cdot\!\hat{\bm p}_2)(\hat{\bm p}_1\!\cdot\!\hat{\bm p}_3)\Bigl[(\hat{\bm p}_1\!\cdot\!\hat{\bm p}_3)^2
\bigl(11(\hat{\bm p}_2\!\cdot\!\hat{\bm p}_3)^2-5\bigr) + 2(\hat{\bm p}_1\!\cdot\!\hat{\bm p}_3)^4 + 2(\hat{\bm p}_2\!\cdot\!\hat{\bm p}_3)^4 - 11(\hat{\bm p}_2\!\cdot\!\hat{\bm p}_3)^2 + 3\Bigr] \nonumber\\
&\quad-2\bigl((\hat{\bm p}_1\!\cdot\!\hat{\bm p}_3)^2-1\bigr)(\hat{\bm p}_2\!\cdot\!\hat{\bm p}_3)
\bigl[2(\hat{\bm p}_1\!\cdot\!\hat{\bm p}_3)^2+2(\hat{\bm p}_2\!\cdot\!\hat{\bm p}_3)^2-1\bigr]\,.
\end{align}

\begin{align}
\mathcal{I}_7(\bm{p}_1,\bm{p}_2,\bm{p}_3) &= 2(\hat{\bm p}_1\!\cdot\!\hat{\bm p}_2)(\hat{\bm p}_1\!\cdot\!\hat{\bm p}_3)^4(\hat{\bm p}_2\!\cdot\!\hat{\bm p}_3)
-4(\hat{\bm p}_1\!\cdot\!\hat{\bm p}_3)^3\Bigl[(\hat{\bm p}_1\!\cdot\!\hat{\bm p}_2)^2\bigl((\hat{\bm p}_2\!\cdot\!\hat{\bm p}_3)^2+1\bigr)
+ (\hat{\bm p}_2\!\cdot\!\hat{\bm p}_3)^2-1\Bigr] \nonumber\\
&\quad+ (\hat{\bm p}_1\!\cdot\!\hat{\bm p}_2)(\hat{\bm p}_1\!\cdot\!\hat{\bm p}_3)^2(\hat{\bm p}_2\!\cdot\!\hat{\bm p}_3)\Bigl[(\hat{\bm p}_1\!\cdot\!\hat{\bm p}_2)^2
\bigl((\hat{\bm p}_2\!\cdot\!\hat{\bm p}_3)^2+11\bigr) + 11\bigl((\hat{\bm p}_2\!\cdot\!\hat{\bm p}_3)^2-1\bigr)\Bigr] \nonumber \\
&\quad -2(\hat{\bm p}_1\!\cdot\!\hat{\bm p}_3)\Bigl[2(\hat{\bm p}_1\!\cdot\!\hat{\bm p}_2)^4\bigl((\hat{\bm p}_2\!\cdot\!\hat{\bm p}_3)^2+1\bigr)
+ (\hat{\bm p}_1\!\cdot\!\hat{\bm p}_2)^2\bigl(2(\hat{\bm p}_2\!\cdot\!\hat{\bm p}_3)^4+(\hat{\bm p}_2\!\cdot\!\hat{\bm p}_3)^2-3\bigr) + 2(\hat{\bm p}_2\!\cdot\!\hat{\bm p}_3)^4-3(\hat{\bm p}_2\!\cdot\!\hat{\bm p}_3)^2+1\Bigr] \nonumber \\
&\quad + (\hat{\bm p}_1\!\cdot\!\hat{\bm p}_2)(\hat{\bm p}_2\!\cdot\!\hat{\bm p}_3)\Bigl[5(\hat{\bm p}_1\!\cdot\!\hat{\bm p}_2)^2
\bigl((\hat{\bm p}_2\!\cdot\!\hat{\bm p}_3)^2-1\bigr) + 2(\hat{\bm p}_1\!\cdot\!\hat{\bm p}_2)^4 + 2(\hat{\bm p}_2\!\cdot\!\hat{\bm p}_3)^4  - 5(\hat{\bm p}_2\!\cdot\!\hat{\bm p}_3)^2 + 3\Bigr]\,.
\end{align}

\begin{align}
\mathcal{I}_8(\bm{p}_1,\bm{p}_2,\bm{p}_3) &= 2(\hat{\bm p}_1\!\cdot\!\hat{\bm p}_2)^5(\hat{\bm p}_1\!\cdot\!\hat{\bm p}_3)(\hat{\bm p}_2\!\cdot\!\hat{\bm p}_3)
-2(\hat{\bm p}_1\!\cdot\!\hat{\bm p}_2)^4\Bigl[2(\hat{\bm p}_1\!\cdot\!\hat{\bm p}_3)^2\bigl((\hat{\bm p}_2\!\cdot\!\hat{\bm p}_3)^2+1\bigr)
+ 2(\hat{\bm p}_2\!\cdot\!\hat{\bm p}_3)^2-1\Bigr] \nonumber \\
&\quad + (\hat{\bm p}_1\!\cdot\!\hat{\bm p}_2)^3(\hat{\bm p}_1\!\cdot\!\hat{\bm p}_3)(\hat{\bm p}_2\!\cdot\!\hat{\bm p}_3) \Bigl[
(\hat{\bm p}_1\!\cdot\!\hat{\bm p}_3)^2\bigl((\hat{\bm p}_2\!\cdot\!\hat{\bm p}_3)^2+11\bigr) + 11(\hat{\bm p}_2\!\cdot\!\hat{\bm p}_3)^2-3 \Bigr]
 \nonumber\\
&\quad-(\hat{\bm p}_1\!\cdot\!\hat{\bm p}_2)^2\Bigl[4(\hat{\bm p}_1\!\cdot\!\hat{\bm p}_3)^4\bigl((\hat{\bm p}_2\!\cdot\!\hat{\bm p}_3)^2+1\bigr)
+ (\hat{\bm p}_1\!\cdot\!\hat{\bm p}_3)^2\bigl(4(\hat{\bm p}_2\!\cdot\!\hat{\bm p}_3)^4+15(\hat{\bm p}_2\!\cdot\!\hat{\bm p}_3)^2-7\bigr)+ 4(\hat{\bm p}_2\!\cdot\!\hat{\bm p}_3)^4-7(\hat{\bm p}_2\!\cdot\!\hat{\bm p}_3)^2+3 \Bigr]
 \nonumber \\
&\quad + (\hat{\bm p}_1\!\cdot\!\hat{\bm p}_2)(\hat{\bm p}_1\!\cdot\!\hat{\bm p}_3)(\hat{\bm p}_2\!\cdot\!\hat{\bm p}_3)\Bigl[(\hat{\bm p}_1\!\cdot\!\hat{\bm p}_3)^2
\bigl(11(\hat{\bm p}_2\!\cdot\!\hat{\bm p}_3)^2-3\bigr) + 2(\hat{\bm p}_1\!\cdot\!\hat{\bm p}_3)^4 + 2(\hat{\bm p}_2\!\cdot\!\hat{\bm p}_3)^4 - 3(\hat{\bm p}_2\!\cdot\!\hat{\bm p}_3)^2 + 1 \Bigr]\nonumber \\
&\quad  + (\hat{\bm p}_1\!\cdot\!\hat{\bm p}_3)^4\bigl(2-4(\hat{\bm p}_2\!\cdot\!\hat{\bm p}_3)^2\bigr) + (\hat{\bm p}_1\!\cdot\!\hat{\bm p}_3)^2\bigl[-4(\hat{\bm p}_2\!\cdot\!\hat{\bm p}_3)^4+7(\hat{\bm p}_2\!\cdot\!\hat{\bm p}_3)^2-3\bigr]
+ 2(\hat{\bm p}_2\!\cdot\!\hat{\bm p}_3)^4 - 3(\hat{\bm p}_2\!\cdot\!\hat{\bm p}_3)^2 + 1\,.
\end{align}
\end{widetext}

\end{document}